\newif\ifanon

\newif\iflipics
\lipicsfalse

\newif\iffull 
\fulltrue

\newif\ifsubmission 
\submissionfalse

\iflipics
    \documentclass[a4paper,UKenglish,autoref,thm-restate,cleveref]{format/lipics-v2021}
\else
    \documentclass[runningheads]{format/llncs}

    \makeatother
    \usepackage[hidelinks, bookmarksopen=true]{hyperref}
    \usepackage{bookmark}
\fi

\usepackage{amsmath}
\usepackage{amssymb}
\usepackage{mathtools}
\usepackage{booktabs}   

\newtheorem{fact}{Fact}

\usepackage{preamble}
\usepackage{tikz}

\newcommand{\mypar}[1]{\smallskip\noindent\textbf{#1.}}

\title{Optimal Reward Allocation via Proportional Splitting\thanks{This is the full version of the paper to appear at the 8th Conference on Advances in Financial Technologies (AFT 2026).}}
\date{\today}

\iflipics
    \ifanon
        \author{Anonymous}{Anonymous}{anonymous}{anonymous}{}
        \authorrunning{Anonymous}
        \Copyright{Anonymous}
    \else
        \author{Lukas Aumayr}{Common Prefix, Greece \and University of Edinburgh, United Kingdom}{lukas.aumayr@gmail.com}{https://orcid.org/0000-0001-8006-3172}{}
        
        \author{Zeta Avarikioti}{Common Prefix, Greece \and TU Wien, Austria}{georgia.avarikioti@tuwien.ac.at}{https://orcid.org/0000-0001-5255-8389}{}
        
        \author{Dimitris Karakostas}{Common Prefix, Greece }{dimitris@commonprefix.com}{https://orcid.org/0000-0002-1868-1719}{}
        
        \author{Karl Kreder}{Dominant Strategies, USA}{karl@dominantstrategies.io}{https://orcid.org/0000-0003-0768-9288}{}
        
        \author{Shreekara Shastry}{Dominant Strategies, USA}{shreekara@dominantstrategies.io}{}{}
        
        \authorrunning{L. Aumayr et al.} 
        
        \Copyright{Lukas Aumayr, Zeta Avarikioti, Dimitris Karakostas, Karl Kreder, and Shreekara Shastry} 

    \fi
\else
    \author{
        Lukas Aumayr\inst{1,2} 
        \and
        Zeta Avarikioti\inst{1,3} 
        \and
        Dimitris Karakostas\inst{1} 
        \and\\
        Karl Kreder\inst{4} 
        \and
        Shreekara Shastry\inst{4} 
    }
    \institute{
         Common Prefix \and University of Edinburgh \and TU Wien \and  Dominant Strategies
    }
    \authorrunning{L. Aumayr, Z. Avarikioti, D. Karakostas, K. Kreder, and S. Shastry}
\fi

\iflipics 


\category{}
\relatedversiondetails[cite={aumayr2025optimalrewardallocationproportional}]{Full Version}{https://arxiv.org/abs/2503.10185}
\funding{This work was partially supported by Quai, Common Prefix,  the Austrian Science Fund (FWF) through the SFB SpyCode project F8512-N, and the Vienna Science and Technology Fund (WWTF) through the projects 10.47379/ICT25056 and 10.47379/ICT22045, and by Input Output (https://iohk.io) through their funding of the Edinburgh Blockchain Technology Lab.}

\supplementdetails[subcategory={Source Code}, cite={evalrepo}, linktext={https://github.com/commonprefix/proportional -reward-splitting-MDP}]{Software}{https://github.com/commonprefix/proportional-reward-splitting-MDP}

\EventEditors{Aggelos Kiayias and Maria Kyropoulou}
\EventNoEds{2}
\EventLongTitle{8th Conference on Advances in Financial Technologies (AFT 2026)}
\EventShortTitle{AFT 2026}
\EventAcronym{AFT}
\EventYear{2026}
\EventDate{October 6--9, 2026}
\EventLocation{London, UK}
\EventLogo{}
\SeriesVolume{395}
\ArticleNo{13}

\fi

\begin{document}

\maketitle

\begin{abstract}
Following the publication of Bitcoin's arguably most famous attack, selfish mining, various works have introduced mechanisms to enhance blockchain systems' game-theoretic resilience.
The only proof-of-work reward rule with a Nash-equilibrium guarantee, FruitChains, demands reward finality on the order of days. The rules that settle in minutes have no such guarantee, and one of them, Reward Splitting, still outperforms FruitChains on most of the metrics that matter in deployment. This paper closes that gap between theory and practice.

We introduce $\protocol$, a two-level transformation for any proof-of-work Nakamoto-style protocol. At the protocol layer, $\protocol$ records low-difficulty samples called \emph{workshares} alongside blocks. At the reward layer, it applies \emph{Proportional Reward Splitting (PRS)}: each height's reward is divided among the competing work objects in proportion to the intrinsic work behind them, with workshares supplying a fresh power estimate at every height. The fork-choice rule and block-production loop are left untouched, so the host chain's security carries over unchanged. Workshares can be discarded once the corresponding rewards mature, leaving zero on-chain footprint.

We prove $\protocol$ is a $\rho$-coalition-safe $\epsilon$-Nash equilibrium for sufficiently large parameters, matching FruitChains in theory. 
To evaluate practical performance, we leverage Markov decision processes and compute the optimal adversarial policy under each utility function, rather than the gain of any one attack. At a six-block confirmation window, $\protocol$ raises the deviation threshold to $38\%$ of mining power and beats every mechanism in that framework on incentive compatibility, subversion gain (except FruitChains above $42$\%), and censorship susceptibility (except FruitChains below $25$\%).
Our construction has been adopted by 
a live PoW blockchain, with sub-kilobyte per-block storage overhead in deployment.
\end{abstract}

\section{Introduction}\label{sec:introduction}
Proof-of-work (PoW) blockchains pay miners to secure the ledger, and the rule that allocates these payments decides whether honest participation is rational.
Selfish mining~\cite{DBLP:journals/cacm/EyalS18} was the first attack to make this precise: by withholding blocks and releasing them strategically, a miner with a minority of the hashing power can capture a share of the rewards exceeding its share of the power. The lasting contribution of that work was the lens rather than the strategy. It analyzed Bitcoin under proportional, rather than absolute, rewards, and showed that under this utility Bitcoin is not an equilibrium. Subsequent work optimized both the attack and its analysis~\cite{DBLP:conf/fc/SapirshteinSZ16,DBLP:conf/ccs/GervaisKWGRC16,DBLP:conf/eurosp/NayakKMS16}. 
Recent events have underscored that these incentives are not merely theoretical.
In August 2025, the Qubic mining pool conducted a documented attack against the PoW-based Monero network, causing multiple chain reorganizations and triggering emergency mitigation proposals from the Monero research community~\cite{qubic2025}.

FruitChains~\cite{DBLP:conf/podc/PassS17} answered selfish mining at the level of mechanism design. It guarantees \emph{fairness}: every miner receives rewards approximately proportional to its power, so deviations are unprofitable. The mechanism rewards low-difficulty objects, called fruits, alongside blocks. Fruits are mined far more frequently than blocks and therefore sample the mining-power distribution more finely, so that rewards track a miner's power rather than the variance in block discovery. FruitChains is a $\rho$-coalition-safe $\epsilon$-Nash equilibrium, the strongest game-theoretic guarantee in this line of work. The guarantee, however, holds only asymptotically. Fairness requires every fruit to be recorded within a window of $R \cdot \kappa$ blocks, with recency parameter $R=17$ and security parameter $\kappa$, which places reward finality on the order of days. 
For any deployment where reward finality must be fast, this is not a usable regime.

The distance between what is provable and what is deployable was made explicit by Zhang and Preneel~\cite{DBLP:conf/sp/0003P19}, who evaluated Nakamoto consensus~\cite{nakamoto2008bitcoin}, FruitChains~\cite{DBLP:conf/podc/PassS17}, Reward Splitting~\cite{DBLP:conf/sp/0003P19,lerner2015decor,camacho2016decor}, and Subchains~\cite{DBLP:journals/ledger/Rizun16} under a single framework. They measured three quantities: (a) incentive compatibility, the fairness of the reward distribution; (b) subversion gain, the profitability of double spending; and (c) censorship susceptibility, the income an attacker can deny honest miners. At realistic parameters, FruitChains was not the best mechanism on most of these metrics. 
\emph{No prior mechanism in this literature simultaneously achieves asymptotic equilibrium guarantees and state-of-the-art practical performance.}

This paper closes that gap. We present \emph{Proportional Reward Splitting (PRS)}, a reward mechanism that is a $\rho$-coalition-safe $\epsilon$-Nash equilibrium for large parameters, matching FruitChains in theory, 
and that achieves state-of-the-art practical performance at a six-block confirmation window, outperforming every mechanism of~\cite{DBLP:conf/sp/0003P19} on incentive compatibility and, with limited exceptions, on subversion gain and censorship susceptibility.

The design begins with separating two roles that existing mechanisms conflate.
On the one hand, Reward Splitting~\cite{DBLP:conf/sp/0003P19,lerner2015decor,camacho2016decor} makes attacks less profitable by compensating non-canonical work, but it uses a coarse $50$--$50$ split whenever competing objects exist at the same height.
On the other hand, FruitChains obtains a finer estimate of mining power from low-work samples, but its theoretical parameters are too large for practical finality.
PRS combines these ideas by using low-work objects, which we call \emph{workshares}, only as a local estimator for proportional splitting: it keeps the anti-forking effect of rewarding competing work, while replacing the coarse split by the measured mining-power distribution. 
For example, an adversary contributing $30\%$ of the power at a height receives approximately $30\%$ of its reward. 
This advantage is largest when the adversary's power is meaningfully below one half and when the workshare sample is accurate; as the adversary approaches one half, the proportional split naturally converges to ordinary Reward Splitting.

PRS could be applied to any PoW Nakamoto-style blockchain (e.g., Bitcoin~\cite{nakamoto2008bitcoin}, PoEM~\cite{poem}, Monero~\cite{cryptonote}) by making two changes: blocks record workshares on-chain, and the reward rule is replaced with proportional splitting. Neither change affects the fork choice rule or block production, so security follows directly from the host protocol's guarantees. 
We call the resulting transformation $\protocol$. We further provide an optimization in which workshares can be kept off-chain, thereby eliminating any on-chain costs of our transformation. Thus, the storage overhead is minimal: workshare data can be discarded once the corresponding rewards finalize.


To evaluate practical performance, we follow the framework of Zhang and Preneel~\cite{DBLP:conf/sp/0003P19}. For each reward mechanism, we construct a Markov decision process (MDP) whose states encode the adversary's private chain, the public chain, the current fork status, and the recent reward history. Solving this MDP yields the optimal adversarial policy for each utility objective, rather than measuring the performance of a fixed attack strategy.

Under this evaluation, \protocol both increases the mining-power threshold beyond which an adversary profits from deviating and lowers its expected gain from doing so. For incentive compatibility, \protocol requires an adversary to exceed $38\%$ of the power to earn more than its fair share, against $35\%$ for the next-best mechanism, Reward Splitting. At $38\%$ power, an adversary earns $38\%$ of all rewards under \protocol, against $40\%$ under Reward Splitting and more than $50\%$ under Bitcoin or FruitChains. 
On subversion gain, 
\protocol matches or beats every alternative across the practical adversarial range, with a strict gap over every other mechanism between $30\%$ and $38\%$ power. On censorship susceptibility, 
it is the best mechanism above $25\%$ power and within two points of the best below. These advantages hold at every workshare and fork-eligibility window we test. 

Finally, we note that the reward mechanism introduced in this paper has been adopted in a deployed system:
a live PoW blockchain 
based on PoEM~\cite{poem},\footnote{\url{https://www.qu.ai}} which records workshares on-chain and allocates rewards proportionally to workshare-estimated mining power. 
Our discussion reports its workshare cap, freshness window, staleness discount, and resulting sub-kilobyte per-block storage overhead. These parameters ground \protocol in a deployed setting and show that the workshares required by \protocol incur only small overhead.


\mypar{Summary of Contributions and Roadmap}
We make the following contributions:
\begin{itemize}
    \item We introduce $\protocol$, a protocol transformation that efficiently incorporates workshares in any PoW Nakamoto-style blockchain without changing the fork choice rule or block production. $\protocol$ uses these workshares to realize PRS, the first reward allocation mechanism for PoW blockchains that is both incentive-compatible and practical for real-world parameters (Section~\ref{sec:protocol}).
    \item We prove that $\protocol$ is a $\rho$-coalition-safe $\epsilon$-Nash equilibrium for sufficiently large parameters, matching the asymptotic guarantee of FruitChains (Section~\ref{sec:analysis}).
    \item We evaluate \protocol via Markov decision processes under practical parameters, comparing against prior mechanisms on incentive compatibility, subversion gain, and censorship susceptibility (Section~\ref{sec:evaluation}).
\end{itemize}


\section{Model and Notation}\label{sec:model-notation}\label{sec:preliminaries}

This section first summarizes the notation (\cref{tab:notation}) and cryptographic primitives 
and  then outlines the execution,
cryptographic, and game-theoretic models of our analysis.

\subsection{Notation and Cryptographic Primitives}
\cref{tab:notation} summarizes the notation used throughout the paper.
Workshare-specific parameters are introduced where they first appear, in
\cref{sec:protocol,sec:analysis}. 

\begin{table}[t]
    \centering
    \caption{Summary of notation.}
    \label{tab:notation}
    \begin{tabular}{@{}ll@{}}
        \toprule
        \textbf{Symbol} & \textbf{Description} \\
        \midrule
        \multicolumn{2}{@{}l}{\emph{Operators and sequences}}\\
        $a.b$ & element $b$ of a (complex) object $a$ \\
        $A \concat A'$ & concatenation of $A$ and $A'$ \\
        $|\cdot|$ & length of an element (list items or bits) \\
        $\ledger \prec \ledger'$ & $\ledger$ is a prefix of $\ledger'$ \\
        \addlinespace
        \multicolumn{2}{@{}l}{\emph{Chains and blocks}}\\
        $\chain[i]$ & $i$-th block of a chain $\chain$ ($\chain[-1]$ is the tip) \\
        $\chain[{:}i],\ \chain[{-}i{:}]$ & first (resp.\ last) $i$ blocks of $\chain$ \\
        $\block_{\tx}$ & block in which transaction $\tx$ is published \\
        $\view$ & execution-trace view \\
        \addlinespace
        \multicolumn{2}{@{}l}{\emph{Parties and power}}\\
        $n$ & total number of parties \\
        $t = \rho n$ & number of adversarial parties \\
        $\rho$ & adversarial mining power, as a fraction of the total \\
        \addlinespace
        \multicolumn{2}{@{}l}{\emph{Mining and timing}}\\
        $\oracleQueries$ & hashing-oracle queries per party per round \\
        $p$ & probability that a player mines a block in a round \\
        $\Delta$ & network delay, in rounds \\
        $w$ & wait-time (liveness) parameter \\
        $k$ & common-prefix (safety) parameter \\
        \addlinespace
        \multicolumn{2}{@{}l}{\emph{Security parameters}}\\
        $\kappa$ & security parameter \\
        $\lambda$ & output length of $\hash(\cdot)$, in bits \\
        \bottomrule
    \end{tabular}
\end{table}

We rely on two cryptographic primitives. The function $\hash(\cdot)$ is a hash
function that maps an arbitrary bitstring to a value of $\secparam$ bits and
models a random oracle.\footnote{$\hash$ should satisfy the standard hash
function properties, namely collision, preimage, and second-preimage
resistance.} The function $\digest(\cdot)$ maps a list of arbitrary bitstrings
to a value of $\secparam$ bits and is collision resistant; it can be
instantiated \eg via a Merkle tree, in which case $\digest(\cdot)$ returns the
tree's root.

\subsection{Execution Model}\label{sec:model}

Our analysis assumes a multi-party setting, following the model of Bitcoin
Backbone~\cite{DBLP:conf/eurocrypt/GarayKL15}. An ``environment'' program
$\env$ drives the execution of a protocol $\proto$, by spawning for each party
$\party$ an instance of an ``interactive Turing machine'' (ITM) which executes
the protocol. Interaction between parties is controlled by a program $C$, such that
$(\env, C)$ create a system of ITMs.  Our analysis applies to ``locally
polynomially bounded'' systems, ensuring polynomial time execution.

\mypar{Network}
The execution proceeds in (a polynomial number of) rounds of size $\Delta$.
In each round $\round$, each party is activated and performs various operations based
on its algorithm. An honest message produced at round $\round$ is delivered to
all other parties at the beginning of round $\round+1$, \ie our analysis assumes
a \emph{synchronous network}.
Finally, we assume a diffuse functionality, \ie a gossip protocol which
allows the parties to share messages without the need of a fully connected graph.

\mypar{Parties}
The set of $\totalParties$ parties $\partySet$ remains fixed for the duration of the
execution. Each party $\party$ can make a fixed number of queries to a hashing oracle, which enables participation in the distributed ledger protocol. 
The adversary controls at most $\adversarialParties = \rho \cdot \totalParties$ parties, out of the total
$\totalParties$ at any given time, and its power is the sum of the corrupted parties' power.

\mypar{Proof-of-Work}
All parties have access to a random oracle $\oracle_\text{rnd}$. Each party can perform
$\oracleQueries$ queries to the oracle in each round,
while the adversary can perform $\adversarialParties \cdot \oracleQueries$ queries per round.
Additionally, the protocol defines
a \emph{difficulty parameter} $p$, 
which denotes the probability that a player successfully creates a block in a round.
In practice, the random
oracle is instantiated as a hash function $\hash(\cdot)$. 
We assume that the codomain of H is large enough such that the outcomes of queries within a batch are practically independent.

\subsection{Cryptographic Model}

In the cryptographic model, some parties are assumed honest (\ie they follow the
protocol) whereas the rest are Byzantine (\ie they act arbitrarily).
Note that incentives do not play a role in the security analysis under this model,
and instead will be considered in the game theoretic analysis.

\mypar{Adversary}
The adversary $\adversary$ is an ITM which, upon activation, may ``corrupt'' a number of
parties by sending a relevant message to the controller after $\env$ instructs it to do so.
Subsequently, when a corrupted party is supposed to be activated, $\adversary$ is activated
instead.
Furthermore, $\adversary$ is ``adaptive'', \ie corrupts parties on the fly,
and ``rushing'', \ie decides its strategy and acts after observing the other
parties' messages.

\mypar{Randomized Execution}
The execution of a protocol is probabilistic, with randomness coming from both honest parties and the adversary, denoted as $A$, which controls all corrupted nodes, as well as the environment $Z$, which provides inputs to honest nodes throughout the protocol's execution. 

We denote a randomly sampled execution trace as $\view \xleftarrow[]{\$} EXEC^\Pi(A, Z, \securityParam)$, where $|\view|$ represents the number of rounds in the execution trace. Specifically, $\view$ is a random variable that captures the joint perspective of all parties, encompassing their inputs, random coins, and received messages (including those from the random oracle). This joint view uniquely determines the entire execution.

\mypar{Restrictions on the Environment}
Constraints on $(A, Z)$~\cite{DBLP:conf/eurocrypt/GarayKL15}. The environment $Z$ and the adversary $A$ must respect certain constraints. We say that a p.p.t. pair $(A, Z)$ is $(n, \rho, \Delta)$-respecting w.r.t. $\Pi$, 
iff for every $\securityParam \in \mathbb{N}$, every view in the support of $EXEC^\Pi(A, Z, \securityParam)$, the following holds:

\begin{itemize}
    \item $Z$ activates $n$ parties in view;
    \item For any message broadcast by an honest player at any time $t$ in view, any player that is honest at
    time $t+\Delta$ or later must have received the message. This means that in the case of newly spawned
    players, it instantly delivers messages that were sent more than $\Delta$ rounds ago. As long as this $\Delta$
    constraint is respected, $A$ is allowed to delay or reorder honest players’ messages arbitrarily.
    \item At every round $r$ in view, $A$ controls at most $\rho \cdot n$ parties.
\end{itemize}

Let $\Gamma(\cdot, \cdot, \cdot)$ be a boolean predicate. We say that a p.p.t. pair $(A, Z)$ is $\Gamma$-compliant w.r.t. protocol
$\pi$ iff:
(i) $(A, Z)$ is $(\totalParties, \rho, \Delta)$-respecting w.r.t. $\pi$;
(ii) $\Gamma(\totalParties, \rho, \Delta) = 1$.




\subsection{Game Theoretic Model}

In the game theoretic model, we assume that all parties are rational.
Specifically, each party $\party$ employs a strategy $\strategy$, which is a set of rules
and actions that the party makes depending on its input. In other words, $\strategy$ defines
the part of the distributed protocol that $\party$ performs. 

Each party has a well-defined utility $\utility$ and chooses a strategy $\strategy$
with the goal of maximizing their utility.
The literature of game theoretic analyses of blockchain protocols considers various
types of utilities~\cite{DBLP:journals/corr/abs-1905-08595}:
(i) absolute rewards;
(ii) absolute profit, \ie absolute rewards minus cost;
(iii) relative rewards, \ie the percentage of the party's reward among all allocated rewards;
(iv) relative profit, \ie relative rewards minus (absolute) cost;
(v) deposits and penalties, \eg using a slashing mechanism;
(vi) external rewards, \eg bribes.

A strategy profile $\strategyProfile$ is a vector of all parties' strategies and
it is an $\epsilon$-Nash equilibrium
if no party can increase its utility more than $\epsilon$ 
by unilaterally changing its strategy 
(Definition~\ref{def:nash-equilibrium}).


\begin{definition}[$\epsilon$-Nash equilibrium]\label{def:nash-equilibrium}
Let $\partySet$ be the set of parties, $\strategySet$ the set of available
strategies, and $\strategyProfile = \langle \strategy_i, \strategy_{-i} \rangle$
a strategy profile, where $\strategy_i$ is the strategy of $\party_i$ and
$\strategy_{-i}$ are the strategies of all other parties. Let
$\utility_i(\strategyProfile)$ denote the utility of $\party_i$ under
$\strategyProfile$. The profile $\strategyProfile$ is an $\epsilon$-Nash
equilibrium with respect to
$\bar{\utility} = \langle \utility_1, \dots, \utility_{|\partySet|} \rangle$ if
$$\forall \party_i \in \partySet \; \forall \strategy'_i \in \strategySet: \quad
\utility_i(\langle \strategy_i, \strategy_{-i} \rangle)
\geq
\utility_i(\langle \strategy'_i, \strategy_{-i} \rangle) - \epsilon.$$
\end{definition}

\iffull
\iffull
\subsection{Property Definitions}\label{sec:background_property_definitions}
\else
\section{Property Definitions}\label{sec:background_property_definitions}
\fi

As our analysis is based on \cite{DBLP:conf/eurocrypt/GarayKL15} and
\cite{PassAnalysis17}, we first recall the definitions of chain growth, chain
quality, and consistency from there. Next, we revisit the fairness
definition of \cite{DBLP:conf/podc/PassS17}.

\subsection*{Chain Growth}

Let:
$$\minchaininc_{t,t'}(\view) = \min_{i,j}|\chaintext^{t+t'}_j(\view)| - |\chaintext^t_i(\view)|$$
$$\maxchaininc_{t,t'}(\view) = \max_{i,j}|\chaintext^{t+t'}_j(\view)| - |\chaintext^t_i(\view)|$$
where we quantify over nodes $i, j$ such that $i$ is honest at round $t$ and $j$ is honest at round $t + t'$ in $\view$.

Let $\growth^{t_0,t1}(\view, \Delta, T) = 1$ iff the following
hold:

\begin{itemize}
    \item \textbf{(consistent length)} for all time steps $t \leq |\view|-\Delta, t+\Delta \leq t' \leq |\view|$, for every two players $i, j$
such that in $\view$ $i$ is honest at $t$ and $j$ is honest at $t'$, we have that $|\chaintext^{t'}_j(\view)| \geq |\chaintext^{t}_i(\view)|$
\item \textbf{(chain growth lower bound)} for every time step $t \leq |\view| - t_0$, we have
$\minchaininc_{t,t_0}(\view) \geq T$.
\item \textbf{(chain growth upper bound)} for every time step $t \leq |\view| - t_1$, we have $\maxchaininc_{t,t_1}(\view) \leq T$.
\end{itemize}

\begin{definition}[Chain growth]
    A blockchain protocol $\Pi$ satisfies $(T_0, g_0, g_1)$-chain growth in $\Gamma$-environments, if for all $\Gamma$-compliant p.p.t. pair $(A, Z)$, there exists some negligible function $\negl$
    such that for every $\securityParam \in \mathbb{N}$, $T \geq T_0$, $t_0 \geq
    \frac{T}{g_0}$
    and $t_1 \leq
    \frac{T}{
    g_1}$
    the following holds:
    \begin{align*}
        Pr[\view \leftarrow EXEC^{\Pi}(A, Z, \securityParam) : growth^{t_0,t_1} (\view, \Delta, T) = 1]\\
        \geq 1 - \negl(\securityParam)
    \end{align*}
\end{definition}

\subsection*{Chain Quality}
The second desired property is that the number of blocks contributed by the adversary is not too large. Given a $\chaintext$, we say that a block $B:=\chaintext[j]$ is honest w.r.t. $\view$ and prefix $\chaintext[:j']$ where $j'<j$ if in $\view$ there exists some node $i$ honest at some time $t \leq |\view|$, such that (i) $\chaintext[:j'] \prec \chain^t_i(\view)$ where $\prec$ denotes ``is a prefix of'' and (ii) $Z$ input $B$ to node $i$ at time $t$. Informally, for an honest node's $\chaintext$, a block $B := \chaintext[j]$ is honest w.r.t. a prefix $\chaintext[:j]$ where $j' < j$, if earlier there is some honest node who received $B$ as input when its local chain contains the prefix $\chaintext[:j']$.

Let $\quality^T(\view,\mu) = 1$ iff for every time $t$ and every player $i$ such that $i$ is honest at $t$ in $\view$, among any consecutive sequence of $T$ blocks $\chaintext[j+1\dots{}j+T] \subseteq \chaintext^t_i(view)$, the fraction of blocks that are honest w.r.t. $\view$ and $\chaintext[:j]$ is at least $\mu$.

\begin{definition}[Chain quality]
    A blockchain protocol $\Pi$ has $(T_0,\mu)$-chain quality, in $\Gamma$-environments if for all $\Gamma$-compliant p.p.t. pair $(A,Z)$, there exists some negligible function $\negl$ such that for every $\securityParam\in\mathbb{N}$ and every $T \geq T_0$ the following holds:
    \begin{align*}
        Pr[\view \leftarrow EXEC^\Pi(A, Z, \securityParam) : \quality^{T}(\view,\mu) = 1]\\
        \geq 1 - \negl(\securityParam)
    \end{align*}
\end{definition}

\subsection*{Consistency}

Let $\consistent^{T}(view) = 1$ iff for all times $t \leq t'$, and all players $i, j$ (potentially the same) such that $i$ is honest at $t$ and $j$ is honest at $t'$ in view, we have that the prefixes of $\chaintext^t_i(view)$ and $\chaintext^{t'}_j(view)$ consisting of the first $\ell = |\chaintext^t_i(view)| - T$ records are identical. This also implies
that the following must be true: $\chaintext^{t'}_j(view) > \ell$, i.e., $\chaintext^{t'}_j(view)$ cannot be too much shorter than $\chaintext^{t}_i(view)$ given that $t' \geq t$.

\begin{definition}[Consistency]\label{def:consistency}
    A blockchain protocol $\Pi$ satisfies $T_0$-consistency in $\Gamma$-environments if for all $\Gamma$-compliant p.p.t. pair (A, Z), there exists some negligible function negl such that for every $\securityParam \in \mathbb{N}$ and every $T \geq T_0$ the following holds:

    \begin{align*}
    Pr\big[\view \leftarrow EXEC^{\Pi}(A, Z, \securityParam) : \consistent^T(\view) = 1\big]\\
    \geq 1 - \negl(\securityParam)
    \end{align*}
\end{definition}


\subsection*{Fairness}

The results from~\cite{PassAnalysis17} are parameterized by the following quantities:

\begin{itemize}
    \item $\alpha:= 1-(1-p)^{(1-\rho)n}$: the probability that some honest player succeeds in mining a block in a round.
    \item $\beta := \rho{}np$: the expected number of blocks an attacker can mine in a round.
    \item $\gamma := \frac{\alpha}{1+\Delta\alpha}$: a discounted version of $\alpha$ which takes into account that messages sent by honest parties can be delayed by up to $\Delta$ rounds and this can lead to honest players redoing work; in essence, $\gamma$ corresponds to the effective mining power.
\end{itemize}

\mypar{Compliant executions for Nakamoto's blockchain} We now specify the compliance predicate 
$\Gamma^{p}_{\mathsf{nak}}$ for Nakamoto's blockchain. 
We say that $\Gamma^{p}_{\mathsf{nak}}=1$ iff there is a constant $\lambda > 1$, such that

$$\alpha(1-2(\Delta+1)\alpha) \geq \lambda\beta$$

where $\alpha$ and $\beta$ are functions of the parameters $n,\rho,\Delta$ and $\securityParam$ as defined above. 
As shown in~\cite{PassAnalysis17}, this condition implies the following:
\begin{fact}
    \label{fact:background1}
    If $(A,Z)$ is a $\Gamma^{p}_{\mathsf{nak}}$-compliant, then $np\Delta < 1$.
\end{fact}

\begin{fact}
    \label{fact:background2}
    If $(A,Z)$ is a $\Gamma^{p}_{\mathsf{nak}}$-compliant, then $\gamma \geq \frac{np}{8}$.
\end{fact}


Subsequently, FruitChains~\cite{DBLP:conf/podc/PassS17} provided a definition for fairness as follows.

Let a player subset selection, $S(\view, r)$, be a function that, given $(\view,
r)$, outputs a subset of the players that are honest at round $r$ in $\view$.
$S$ is a $\phi$-fraction player subset selection if $S(\view, r)$ always outputs a set of size $\phi n$ (rounded upwards) where $n$ is the number of players in $\view$.

Given a player subset selection $S$, a \emph{record $m$ is $S$-compatible w.r.t. $\view$ and prefix
    $\chaintext$} if there exists a player $j$ and round $r'$ such that $j$ is in $S(\view, r')$, the environment provided $m$ as an input to $j$ at round $r'$, and $\chaintext \prec \chaintext^{r'}_j(\view)$, where $\prec$ denotes ``is a prefix of''.

Let $\quality^{T,S}(\view, \mu) = 1$ iff for every round $r$ and every player $i$ such that $i$ is honest in round $r$ of $\view$, we have that among any consecutive sequence of $T$ records $\chaintext^{r}_i(\view)[j + 1 : j + T]$, the fraction of records that are $S$-compatible w.r.t. $\view$ and prefix $\chaintext^{r}_i(view)[: j]$ is at least $\mu$.

\begin{definition}[Fairness]
    A blockchain protocol $\Pi$ has (approximate) fairness $(T_0, \delta)$ in $\Gamma$-environments, if
    for all $\Gamma$-compliant p.p.t. $(A, Z)$, every positive constant $\phi \leq 1-\rho$, every $\phi$-fraction subset selection
    $S$, there exists some negligible function $\epsilon$ such that for every $\securityParam \in \mathbb{N}$ and every $T \geq T_0$ the following holds:
    \begin{align*}
    Pr[\view \leftarrow EXEC^\Pi(A, Z, \securityParam) : \quality^{T,S}(\view,(1 - \delta)\phi)) = 1]\\
    \geq 1 - \epsilon(\securityParam)
    \end{align*}
\end{definition}

\fi
\section{The \protocol Protocol}\label{sec:protocol}
\protocol is a two-level transformation on any proof-of-work longest-chain protocol.
At the \emph{protocol level}, it adds the recording of low-work objects, called
\emph{workshares}, on-chain alongside blocks. At the \emph{reward level}, it
replaces the existing allocation rule with proportional splitting: each height's
reward is distributed among all competing objects at that height in proportion
to their intrinsic work. Neither modification touches the host protocol's
fork-choice rule or block-production mechanism; as such, \protocol inherits the host protocol's security guarantees.



\mypar{$\protocol$ Overview}
The intrinsic work of any hash-based object $\mathsf{obj}$ is
\begin{equation}\label{eq:intrinsic-work}
    \work(\mathsf{obj}) = \secparam - \lfloor \log \hash(\mathsf{obj}) \rfloor,
\end{equation}
the number of leading zero bits in its hash: the more zero bits, the higher the
work. $\protocol$ uses this measure to
estimate each miner's relative power via workshares. A workshare has the same
structure as a block and is produced by the same mining loop, but satisfies a
lower work threshold. 
Parties propagate workshares through the
network and publish them in blocks, as they do with transactions. Since
workshares and blocks are outputs of the same hash query, a miner neither
chooses between the two nor knows in advance whether a query yields a workshare,
a block, or neither; the hash function models a random oracle. Recording workshares is the only modification
$\protocol$ makes to the host protocol. Because workshares affect neither
fork choice nor block production, $\protocol$ inherits the host protocol's
security properties unchanged.

At a high level, in each round a miner runs the
PoW loop; a successful query yields a workshare, a block, or both, depending on
the leading zero bits of its hash. The miner broadcasts every new object, and
whoever mines the next block records the workshares and uncles it has seen
inside that block. Once the relevant blocks are stable, the work recorded for a
given height, across its block and the workshares pointing to it, determines how
that height's reward is split among miners (\cref{sec:rewards}). {Workshares thus
serve one purpose: they sample mining power between blocks, so that rewards
track power closely. }

The accuracy of this power estimate is not a fixed per-block quantity: it
depends jointly on the workshare rate, the block-production rate, and the
reward-maturity window. A deployment can improve accuracy by increasing the
workshare rate, by extending the maturity window, or both, without requiring all
samples to appear in a single block.

In the remainder of this section, we describe the protocol's data objects,
its core algorithm, the chain-selection and ledger-extraction mechanisms,
the reward allocation rule, and an optimization that removes the on-chain
storage overhead.

\subsection{Blocks and Workshares}\label{subsec:transactions-blocks}

A work object in $\protocol$ has the following format:
$$
\workObject = \langle (\digest(\txList); \digest(\shareList); h_{\block'}; h_{\hat{\block}}), \nonce \rangle
$$
where $\txList$ is the transaction list, $\shareList$ is the list of work
objects (workshares or uncles) published when $\workObject$ acts as a
block,\footnote{An uncle satisfies the block work threshold but is not part of
the canonical chain, i.e., it has been forked out.} $h_{\block'}$ and
$h_{\hat{\block}}$ reference two prior blocks via their hashes, and $\nonce$ is
the PoW nonce. A full block is the work object together with the explicit lists
$\txList$ and $\shareList$. When an object is interpreted as a workshare,
$\digest(\txList)$ and $\digest(\shareList)$ are ignored; those fields are used
only by the chain-selection and ledger-extraction mechanisms. Let $\recencyParam$ denote the recency window: a workshare or uncle is eligible for inclusion only if the block it references is at most $\recencyParam$ blocks behind the including block.


In more detail, the validity rules for blocks and workshares are as follows.
In addition, we assume the existence of a transaction validity predicate, which
specifies the application logic of the ledger, \eg preventing double spending.

\mypar{Block validity}
$\block = (\langle (\digest(\txList); \digest(\shareList); h_{\block'}; h_{\hat{\block}}), \nonce \rangle, \txList)$
is valid \wrt a chain $\chain$ if the following conditions hold:
\begin{itemize}
    \item $\digest(\txList)$ is the digest of $\txList$.
    \item $h_{\block'}, h_{\hat{\block}}$ are references to $\chain[-1], \chain[-k]$ respectively.
    \item Every \wo (workshare or uncle) in $\shareList$ is valid \wrt $\chain$ (see below);
    \item No \wo is present in $\chain$ multiple times.
    \item Every transaction in $\txList$ satisfies the protocol's validity
        predicate.
    \item $\work(\block) > \powBlockThreshold$, where $\powBlockThreshold$ is a
        protocol parameter that defines the PoW lower bound for a block to be valid.
\end{itemize}

\mypar{Uncle validity}
$\block = (\langle (\digest(\txList); \digest(\shareList); h_{\block'}; h_{\hat{\block}}), \nonce \rangle, \txList)$ is a valid uncle \wrt chain $\chain$, if
all of the following hold:
\begin{itemize}
    \item $\block$ is included in $\shareList$ of exactly one block $\overline{\block}$ in $\chain$ and $\block \notin \chain$;
    \item $h_{\block'}$ is a reference to $\chain'[-1]$, which in turn is either a valid block in $\chain$ or a valid uncle \wrt $\chain$;
    \item $h_{\hat{\block}}$ is a reference to $\chain'[-k]$, which is a valid block in $\chain$;
    \item Let $\overline{\chain}$ be the prefix of $\chain$ that goes up to this block $\overline{\block}$. 
    $|\chain'|$ is between $|\overline{\chain}| - \recencyParam$ and $|\overline{\chain}| - 1$;
    \item Following the reference $h_{\block'}$ to a block $\block'$, and then this block's reference and so on at most $\safetyParam$ times, 
    gives a block that is part of $\chain$;
    \item $\work(\block) > \powBlockThreshold$.
\end{itemize}

\mypar{Workshare validity}
$\share = \langle (\digest(\txList); \digest(\shareList); h_{\block'}; h_{\hat{\block}}), \nonce \rangle$ is valid \wrt chain $\chain$ if:
\begin{itemize}
    \item $\share$ is included in $\shareList$ of exactly one block $\overline{\block}$ in $\chain$;
    \item  Let $\overline{\chain}$ be the prefix of $\chain$ that goes up to this block $\overline{\block}$. 
    $h_{\block'}$ is a reference to any valid block or uncle with a height between $|\overline{\chain}| - \recencyParam$ and $|\overline{\chain}| - 1$;
    \item $\work(\share) > \powShareThreshold$, where $\powShareThreshold$ is a
        protocol parameter that defines the PoW lower bound for a workshare to be valid.
\end{itemize}

Intuitively, $\powShareThreshold$ and $\powBlockThreshold$ define the minimum
amount of intrinsic work that a workshare or block should have, respectively, and it
holds that $\powShareThreshold \leq \powBlockThreshold$.

\mypar{Chain validity}
Finally, a chain $\chain$ is valid \wrt a genesis block $\genesis$ (denoted
$\validate_\genesis(\chain)$) if every block in $\chain$ is valid and they form
a chain that starts from $\genesis$.

\subsection{Main Protocol}\label{subsec:main-protocol}

The main function of $\protocol$ is defined in \cref{alg.protocol}. 
First, the party is constructed using the \constructor function (Line~\ref{poem2.alg-backbone.constructor}).
In every round, each party is executed by the environment using the function \execute
(this function is due to the round-based nature of our time model).
The algorithm maintains a local chain $\chain$, which is initialized to the genesis
block $\genesis$, and proceeds as follows.

\begin{algorithm}[htb]
    \caption{\label{alg.protocol} The $\protocol$ protocol}
    \begin{algorithmic}[1]
        \Statex
        \Let{\genesis}{\emptyset}
        \Let{\chain}{[\,]}
        \Function{$\constructor$}{$\genesis'$}\label{poem2.alg-backbone.constructor}
            \State $\genesis \leftarrow \genesis'$ \Comment{Select Genesis Block}
            \State $ \chain \leftarrow [\genesis]$ \Comment{Add Genesis Block to start of chain}
            \State \textsf{round} $\leftarrow 1$
        \EndFunction
        \Function{$\execute_{\digest,\hash}$}{$1^\securityParam$}
            \State $\chainSet \leftarrow \receive()$\label{poem2.receive-chains} \Comment{Receive chains from the network}
            \State $\chain \leftarrow \maxvalid(\{ \chain \} \cup \chainSet)$ \Comment{Apply host protocol's chain-selection rule} 
            \Let{\langle \txList, \shareList \rangle}{\textsc{input}()}\label{poem2.receive-txs}
            \Let{\blockList}{\emptyset}
            \State{$\nonce \getsrandomly \{0,1\}^\secparam$}
            \For{$i \gets 1 \text{ to } \oracleQueries$}
                \Let{\workObject}{\langle (\digest(\txList); \digest(\shareList); \hash(\chain[-1]); \hash(\chain[-\safetyParam])), \nonce \rangle}
                \If{$\work(\workObject) > \powShareThreshold$}
                    \Let{\shareList}{\shareList \concat \workObject}
                \EndIf
                \If{$\work(\workObject) > \powBlockThreshold$}
                    \Let{\block}{\langle \workObject, \txList \rangle}
                    \Let{\blockList}{\blockList \concat \block}
                    \Let{\chain}{\chain \concat \block}
                \EndIf
                \Let{\nonce}{\nonce + 1}
            \EndFor
            \State $\diffuse(\chain, \shareList\concat \blockList)$
            \State {\textsf{round}} $\leftarrow$ {\textsf{round}+1}
        \EndFunction
    \end{algorithmic}
\end{algorithm}

\mypar{Chains}
The party first receives from the network the set of available chains via the function $\textsc{receive}$
(Line~\ref{poem2.receive-chains}). Out of all the available chains, including the local chain $\chain$, the party
adopts the chain selected by the host protocol's chain-selection rule, denoted
$\maxvalid$.

\mypar{Mempool}
The party receives the set of transactions in the mempool
$\txList$ via the function $\textsc{input}$ (Line~\ref{poem2.receive-txs}) and
attempts to publish them in the newly-mined block. Specifically, the party
orders the transactions in $\txList$ in an arbitrary manner in order to form the
list $\txList$; for ease of notation, we abstract the ordering process outside
of the protocol of~\cref{alg.protocol}. The list's digest
$\digest(\txList)$ is included in the new block's header.

\mypar{Workshares}
The party receives from the network workshares $\shareList$, including 
uncle blocks 
(Line~\ref{poem2.receive-txs}), and attempts to publish them in the
new block. As opposed to transactions, the order of
workshares does not matter, since they are only used for reward allocation (cf.
\cref{sec:rewards}).

We note that $\txList$ and $\shareList$ are both sanitized to contain valid
objects \wrt the party's local chain $\chain$, so that the transactions and
workshares can be published in a new block. For ease of notation we omit this
process from~\cref{alg.protocol} and assume that $\txList$ and $\shareList$
(obtained from $\textsc{input}$) are sanitized \wrt the local chain
$\chain$ at all rounds. In particular, all workshares fulfill the PoW inequality, 
there are no duplicate workshares in the chain, and uncle blocks are not part
of the chain.

\mypar{Proof-of-Work loop}
The main part of the protocol is the PoW loop, during which the party \emph{potentially} creates
workshares and blocks. Specifically, on each round the party can
execute $\oracleQueries$ queries to the hash function. For each query, the work object
includes a PoW nonce $\nonce$ and, depending on the amount of work that the
object has, it may be a valid workshare and/or block. If a workshare is created, then it
is immediately inserted into the list $\shareList$. This has the implication that
a workshare which is created on some round $r$ may be published on a block that is
created on the same round $r$. Similarly, if a block is created, it is
immediately appended to the local chain $\chain$. Regardless of whether a workshare or
block is created, the party exhausts all available hashing queries at their
disposal without interrupting the PoW loop. Therefore, a party that follows the
protocol may create multiple workshares and/or blocks on a single round.

Finally, the party diffuses to the network their local chain and list of
workshares and proceeds to the next round.

\subsection{Chain Selection and Ledger Extraction}
\label{subsec:chain-selection}

Chain selection follows the host protocol's $\maxvalid$ rule.
Ledger extraction\ifsubmission\else{} (\cref{alg.extract})\fi{} is parameterized
by the safety parameter $\safetyParam$. Given a chain $\chain$, the last
$\safetyParam$ blocks are excluded from the ledger extraction process. The
remaining blocks in $\chain$ are parsed sequentially, starting from the genesis
block $\genesis$, and the ledger $\ledger$ is formed by concatenating each
transaction in the list $\txList$ of each block.





\subsection{Proportional Reward Splitting Mechanism}\label{sec:rewards}

We now specify the reward rule of $\protocol$. At each block height, PRS
distributes a fixed reward among all eligible work objects at that height in
proportion to their intrinsic work. This refines Reward
Splitting~\cite{DBLP:conf/sp/0003P19,lerner2015decor,camacho2016decor}, which
also distributes a height's reward across competing objects but uses a fixed
$50$-$50$ split regardless of the work behind them. PRS replaces this coarse
split with the measured power distribution: a miner contributing $30\%$ of
the work at a height receives $30\%$ of that height's reward.\footnote{The
power distribution is not directly observable; PRS estimates it from
workshares, as analyzed in \cref{sec:evaluation}.} The estimate is recomputed
at every height to track fluctuations in mining power as they occur.

\mypar{Eligibility windows}
Two parameters control which objects are eligible for rewards at a given
height (\cref{fig:rewards}). The \emph{recency window} $R$ sets the
finalization delay: the reward for height $h$ is settled at height $h + R$,
and only work objects published within that window are counted. The
\emph{fork eligibility window} $\safety$ limits which forks contribute: an
object is eligible only if the block it references lies at most $\safety$
blocks into a fork, excluding objects that build on deeply invalid chains.
In our analysis, $\safety$ is set to the common-prefix (consistency) parameter $T_0$ of the
host protocol\ifsubmission\else{} (cf. \cref{def:consistency})\fi.

\mypar{Reward rule}
At height $h + R$, each eligible work object at height $h$ receives a
fraction of the height's reward equal to its intrinsic work divided by the
total intrinsic work of all eligible objects at that height. The height of a
workshare or uncle is the height of the block it references via
$h_{\block'}$, not the block in which it is published. Maturity delays of
this kind are standard: Bitcoin requires $100$ blocks before a coinbase
transaction can be
spent.\footnote{\url{https://learnmeabitcoin.com/technical/mining/coinbase-transaction/\#coinbase-maturity}}

\begin{figure}[ht]
    \centering
    \resizebox{0.65\columnwidth}{!}{
        \input{figures/fig_uncle_fruit}
    }
    \caption{Uncles can be referenced up to $R$ blocks away. Objects pointing
    to blocks more than $\safety$ blocks into a fork are ineligible. At
    height $h_i + R$, the reward at height $h_i$ is distributed
    proportionally to the intrinsic work of the eligible objects at $h_i$.}
    \label{fig:rewards}
\end{figure}

\subsection{Removing on-chain storage overhead}\label{sec:removestorage}
\protocol does not modify the $\maxvalid$ fork-choice rule of the underlying proof-of-work chain, so work objects play no role in chain selection, only in reward allocation. Instead of recording them on-chain within blocks, it suffices that miners share them over the P2P network. Miners will verify reward (coinbase) transactions based on the work objects sent along with the block.
Once the rewards for height $h_i$ are finalized at height $h_i + R$ (cf. \cref{fig:rewards}), all work objects older than $R + k$ blocks behind the current tip can be safely discarded. Bootstrapping nodes need only download the longest chain; they can skip work object verification entirely. Thus, the storage overhead of \protocol is bounded by the work objects of the last $R + k$ blocks. 

This storage improvement comes with a small tradeoff: a bootstrapping node cannot independently recompute past reward allocations. However, under the honest-majority assumption, honest miners validate reward allocations before extending a block, so a bootstrapping node can still infer their historical validity by following the heaviest PoW chain.


\section{Theoretical Analysis}\label{sec:analysis}

Our analysis closely follows that of FruitChains~\cite{DBLP:conf/podc/PassS17}. The protocol difference mainly affects the
analysis of workshare freshness. In the rest of this section, we prove
workobject freshness, consistency, growth, and fairness, ultimately leading to
incentive compatibility. We denote the probability that any player mines a
block in a round as $p$, and similarly the probability that a player mines a
workshare in a round as $p_f$, such that $q = \frac{p_f}{p}$.

\subsection{Main Theorem}

Theorem~\ref{thm:main} is the main theoretical result of our work, which shows
that the workshare-based protocol guarantees the useful properties that we aim
it to achieve, namely consistency, chain growth, and fairness.

\begin{theorem}[Security of \sys]\label{thm:main} 
    For any constant $0 < \delta < 1$, and any $p, p_f$, let $R = 16$,
    $\safetyfruit = 2qR\safety$, and $T_0 = 5\frac{\safetyfruit}{\delta}$. Then, in $\Gamma_{\sys}^{p,p_f,R}$-environments, the Workshare protocol denoted $\Pi_{\sys}(p, p_f, R)$ satisfies
    (i)~$\safetyfruit$-\emph{consistency},
    (ii)~\emph{chain growth rate} $(T_0, g_0, g_1)$ with
    $g_0 = (1 - \delta)(1 - \rho)np_f$ and $g_1 = (1 + \delta)np_f$, and
    (iii)~\emph{fairness} $(T_0, \delta)$.
\end{theorem}



\iffull
\subsection{\wo Freshness}


A key property is for any \wo (or \emph{fruit} to follow \cite{DBLP:conf/podc/PassS17}) mined by an honest player to stay sufficiently fresh to be incorporated. In this section, we prove freshness. Note that we achieve a slightly better freshness parameter ($R=16$) compared to FruitChains~\cite{DBLP:conf/podc/PassS17} ($R=17$).

Let $\unclefreshness(\view, \liveness, \safety) = 1$ iff for every honest player $i$  and for every round $r < |view| - \liveness$, if $i$ mines a \wo at $r$ in view, then for every honest player $j$ there exists an index $pos$, such that the \wo is at $pos$ in the record chain (w.r.t. Nakamoto's protocol) of $j$ at every round $r' \geq r + \liveness$, and $pos$ is at least $\safety$ blocks away from the end of the chain.

Let

$$wait = 2\Delta+\frac{2\safety}{\gamma}$$

\begin{lemma}
    Let $R=16$. For any $p$, $p_f$, for any $\Gamma_{\sys}^{p,p_f,R}$-compliant $(A,Z)$, there exists a negligible function $\epsilon$ such that for any $\securityParam \in \mathbb{N}$,

$$Pr\big[\view \leftarrow EXEC^{\Pi_{\sys}%
(p,p_f,R)}(A, Z, \securityParam) :$$ $$ \unclefreshness(\view,wait,\safety) = 1\big]
\geq 1 - \epsilon(\securityParam)$$
    
\end{lemma}

\begin{proof} Disregard the blockchain consistency, 
liveness, and chain growth failure events, which only happen with negligible probability. Let $wait = wait(\safety, n, \rho, \Delta)$. 

    \begin{itemize}
        \item By \emph{blockchain consistency}, at any point in the execution, whenever an honest player mines a \wo (i.e., a block, an uncle block or a workshare) $f$, the block pointed to by that \wo is at some \emph{fixed} position $pos$ on the blockchain of every honest player, now and at every time in the future. 
        \item Note that if $\work(\wo) > \powBlockThreshold$, it is not clear at the time of mining whether it will be a block or an uncle block.
        \item Honest players try to mine \wos that point to the last block in their chain. 
        \item Let $\ell$ denote the length of the chain of the player that mines $f$; by definition $pos = \ell$
        \item By the description of the protocol, if the \wo $f$ is mined at round $r'$, it gets seen by all honest players by round $r'+\Delta$; additionally, when this happens, all honest players attempt to add $f$ to their chain as long as it remains recent (w.r.t. all honest players).
        \item By \emph{liveness}, the \wo gets thus referenced by a block that is in the record chain of all honest players at some position $pos$ that is at least $\safety$ blocks away from the end of the chain by round 
        $$r' + \Delta + (1+\delta)\frac{\safety}{\gamma} \leq r' + wait- \Delta$$
        \item By the upper bound on chain growth, at most $$(1+\delta)np(\Delta+\frac{2\safety}{\gamma})$$ blocks are added in time $wait-\Delta$. Thus, by round $r' + wait - \Delta$, no honest player has ever had a chain of length $\ell'$, such that 
        
        $$\ell' > \ell + (1+\delta)np(\Delta+\frac{2\safety}{\gamma})$$

        \item Thus, by round $r' + wait - \Delta$, for every such honest player's chain length $\ell'$, we have 
        
        $$pos = \ell \geq \ell' - (1+\delta)np(\Delta+\frac{2\safety}{\gamma})$$
        
        \item By our compliance assumption and by Fact \ref{fact:background1} and Fact \ref{fact:background2}, we have that $\gamma \geq \frac{np}{8}$ and $np\Delta < 1$, thus 

        $$pos > \ell'-(1+\delta)-(1+\delta)16\safety \geq \ell' - 16\safety = \ell' - R\safety$$
        \item By \emph{consistency}, all honest players agree that $f$ is found at position $pos$ in their blockchain at any point after $r'+wait-\Delta$; additionally, by the \emph{consistent length property} all honest players agree that position $pos$ is at least $\safety$ from the end of the chain by $r'+wait-\Delta+\Delta = r'+wait$.
        \item It remains to show that the block $\block$ that is referenced by $f$ is either a valid block of the chain $\chain$ of honest users or a valid uncle, as this is the condition for \wos to be rewarded. Note that honest parties will reference the last block they see in their chain, which is valid, but not necessarily stable. Should it happen that the block does not end up in $\chain$, it thus has to become an uncle. Suppose there exists another block $\block'$ that is competing with $\block$. In any block that is building on top of $\block'$, $\block$ is treated as a \wo (uncle). We have already proven that \wos get incorporated into the chain of every honest player; thus, $\block$ will be incorporated into the chain of every honest player.
    \end{itemize}
\end{proof}

    We also observe the following fact about $wait$, which says that the expected number of \wo{}s mined by all players during $wait + 2$ is upper bounded by $\safetyfruit$.

    \begin{fact}
    \label{fact1}
        For any $p$, $p_f$, any $\Gamma_{\sys}^{p,p_f,R}$-compliant $(A,Z)$,
        $$(wait+2)\cdot np_f\leq\safetyfruit$$
    \end{fact}
    \begin{proof}
        Note that by Fact \ref{fact:background1} and Fact \ref{fact:background2}, we have that $\gamma > \frac{np}{8}$ and $np\Delta<1$, thus

        $$(wait+2)\cdot np_f = (2\Delta+2\frac{\safety}{\gamma}+2)\cdot{}qpn\leq{}2q+2\safety\cdot{}8q+2\leq{}2qR\safety=\safetyfruit$$
    \end{proof}

\subsection{Some Simplifying Assumptions}
\label{sec:simplif-assump}

Towards proving our main theorem, we state some simplifying assumptions that can be made without loss of generality. These assumptions (which all follow from properties of the random oracle $H$) will prove helpful in our subsequent analysis.

\begin{itemize}
    \item \textbf{WLOG1:} We may without loss of generality assume that honest players never query the RO on the same input. More precisely, we analyze an experiment where if some honest player wants to query it on an ``old'' input, it re-samples the nonce until the input is ``new''; since the nonce is selected from $\{0, 1\}^\securityParam$, this ``re-sampling'' experiment is identical to the real one except with negligible probability, thus we can WLOG analyze it.
    \item \textbf{WLOG2:} We may, without loss of generality, assume that any \wo{} that points to a block $b$ which was first mined at time $t$ has been mined after $t$. Additionally, any \wo{} that points to a block that comes after $b$ in a valid chain must have been mined after $t$. (If not, we can predict the outcome of the random oracle $H$ on some input before having queried $H$, which is a contradiction. We omit the standard details.)
    \item \textbf{WLOG3:} We may assume without loss of generality that all \wo{}s mined by honest players are ``new'' (i.e., different from all valid \wo{}s previously seen by honest players); this follows from
WLOG1 and the fact that the probability of seeing a collision in the random oracle is negligible (by a simple union bound over the number of random oracle queries).
    \item \textbf{WLOG4:} We may assume without loss of generality that any valid \wo{} which appears in some honest player's chain at round $r$ was mined before $r$; this follows from the unpredictability of the
random oracle (and a simple union bound over the number of random oracle queries).
    \item \textbf{WLOG5:} We may assume without loss of generality that there are no ``blockchain collisions'', namely, there are no two \emph{different} valid sequences of blocks which end with the same block. 
\end{itemize}

We now turn to proving the three security properties.

\subsection{\wo Consistency}\label{subsec:consistency}

Disregard chain growth, consistency, and chain quality failure events--they happen with negligible probability. 
Consider some view $\view$ in the support of $EXEC^{\Pi_{\sys}(p,p_f,R)}(A, Z, \securityParam)$, rounds $r, r'$, s.t. $r' \geq r$, and players $i,j$ that are honest respectively at $r, r'$ in view. By \emph{consistency}, the chains of $i,j$ at $r,r'$ agree except for potentially the last $\safety$ blocks in the chain of $i$. Let $C = b_0, \dots, b_{|C|}$ denote those blocks on which they agree and let $b_{|C|+1},\dots$ denote the maximum $\safety$ blocks in the chain of $i$ at $r$ which are not in the chain of $j$ at $r'$. We now bound the number of \wos that can be referenced in these remaining (maximum $\safety$) inconsistent blocks.

\begin{itemize}
    \item By the ``recency condition'' of valid \wo{}s, any valid \wo in the chain of $i$ at $r$ which is after $C$ must point to a block $b_{j'}$ such that $j' > |C|-R\safety$.
    \item By the \emph{chain quality condition}, there exists some $j''$ s.t. $|C|-R_\safety-\safety\leq j'' \leq|C|-R_\safety$ and $b_{j''}$ was mined by an honest player. Let $r'_0$ denote the round when this block was mined.
    \item Note that $r'_0$ was mined by an honest player holding a chain of length $j'' \geq |C|-R_\safety-\safety$; additionally, at $r$, $i$ is honest, holding a chain of length at most $|C| + \safety$ (recall that |C| contains the blocks on which $i$ and $j$ agree, and by consistency, all but the last $\safety$ blocks in the chain of $i$ must be in the chain of $j$). Thus, by the \emph{chain growth upper bound}, at most
    $$\mu = (1+\delta)\frac{2\safety+R\safety}{np}$$
    rounds could thus have elapsed between $r'_0$ and r.
    \item By WLOG2, any \wo{} which gets added after $C$ must have been mined after $r'_0$. By WLOG4, any such \wo{} that is part of the chain of $i$ by $r$ was mined before $r$.
    \item We thus conclude by the Chernoff bound that for every sufficiently small $\delta'$ except with probability $e^{=\Omega(np_f\cdot\frac{\safety(R+2}{np})} = e^{-\Omega(q(R+2)\safety})$, there were at most 
    $$(1+\delta')^2\cdot{}np_f\cdot{}\frac{\safety(R+2)}{np}=(1+\delta')^2q(R+2)\safety<2qR\safety=\safetyfruit$$
    ``inconsistent'' \wo{}s in the chain of $i$ at $r$
\end{itemize}

\subsection{\wo Growth}

\mypar{Consistent length} The consistent length property follows directly from the consistent length
property of the underlying blockchain.

\mypar{Lower bound} Consider any $r, t$ and players $i, j$ that are honest respectively at round $r$ and $r + t$.
Consider the $t$ rounds starting from round $r$.

\begin{itemize}
    \item By the \wo{} freshness condition, every \wo{} that is mined by some honest 
    party by round $r + t - wait$ gets 
    incorporated into (and remains in) the 
    chain of player $j$ by $r + t$.
    \item By the Chernoff bound, in the $t - wait$ rounds from $r$ to $r + t - wait$, except with probability $e^{-\Omega((t-wait)\alpha_f)}$, the honest parties mine at least
    $$(1 - \delta')(t-wait)\alpha_f$$
    \wo{}s (where $\delta'$ is some arbitrarily small constant), which are all included in the chain of $j$ at $r + t$. Additionally, by WLOG3 they are all “new” (i.e., not included in the chains of $i$ at $r$) and different.
    \item Finally, by \wo{} consistency (proved in \cref{subsec:consistency}), we have that all but potentially $\safetyfruit$ of the \wo{}s in the chain of $i$ at $r$ are still in the chain of $j$ at $r + t$.
    \item We conclude that, except with probability $e^{-\Omega((t-wait)\alpha_f)}$ the chain of $j$ at $r+t$ contains at least
    $$(1-\delta')(t-wait)\alpha_f-\safety_f$$
    more \wo{}s than the chain of $j$ at $r$. By Fact \ref{fact1}, $wait\cdot\alpha_f=wait \cdot(1-\rho)np_f\leq (1-\rho)\safetyfruit\leq\safetyfruit$; thus, we have at least
    
    \begin{equation}
    \label{eq:fruitgrowthlower}
        (1-\delta)(t-wait)\alpha_f-\safetyfruit\geq(1-\delta)\alpha_f{}t-2\safetyfruit
    \end{equation}
    new \wo{}.
\end{itemize}

    We conclude by noting that this implies that a \wo{} growth lowerbound of $g_0 = \frac{1}{1+\delta}\alpha_f \geq (1-\delta)\alpha_f$ in the desired regime: Consider any $T\geq \frac{5\safetyfruit}{\delta}$ and any 
    $$t\geq \frac{T}{g_0}=\frac{T}{\frac{\alpha_f}{1+\delta}}$$
    
    As shown above (see \cref{eq:fruitgrowthlower}), except with a probability of $e^{-\Omega((t-wait)\alpha_f)}$ the chain must have grown by at least 
    $$T(1+\delta)(1-\delta')-2\safetyfruit=T(1+\frac{\delta}{2})(1-\delta')+T\frac{\delta}{2}(1-\delta')-2\safetyfruit$$

    For a sufficiently small $\delta'$, the first term is greater than $T$, and the second term is greater than $2\safetyfruit$, and thus the chain must have grown by at least $T$. Finally note that by \cref{eq:fruitgrowthlower}

    $$e^{-\Omega((t-wait)\alpha_f)}=e^{-\Omega(t\alpha_f-\safetyfruit)}=e^{-\Omega(T-\safetyfruit)}=e^{-\Omega(5\safetyfruit-\safetyfruit)}=e^{-\Omega(\safety)}$$
    Thus, the chain growth is guaranteed except with negligible probability.

\mypar{Upper bound} Disregard the chain growth, consistency, and chain quality failure events--they happen with negligible probability. Consider some view  $\view$ in the support of $EXEC^{\Pi_{\sys}(p,p_f,R)}(A, Z, \securityParam)$, rounds $r, r'$, s.t. $r' = r + t$, and players $i,j$ that are honest respectively at $r, r'$ in view. By \emph{consistency}, the chains of $i,j$ at $r,r'$ agree except for potentially the last $\safety$ blocks in the chain of $i$. Let $C = b_0, \dots, b_{|C|}$ denote those blocks on which they agree and let $b_{|C|+1},\dots$ denote the blocks in the chain of $j$ at $r'$ which are not in the chain of $i$ at $r$ (there may be more than $\safety$ such blocks since we are looking at the chain of $j$ at a later time $r'$); We now upper bound the number of \wo{}s in the new blocks in the chain of $j$ which come after $C$, similarly to the \wo{} consistency proof (the main difference is that we now consider the chain of $j$ as opposed to the chain of $i$). The details follow:

\begin{itemize}
    \item By the ``recency condition'' of valid \wo{}, any valid \wo{} in the chain of $j$ at $r'$ which is after $C$ must point back to a block $b_{j'}$, such that $j' > |C|-R\safety$.
    \item By the \emph{chain quality condition}, there exists some $j''$ s.t. $|C|-R\safety-\safety\leq{}j''\leq|C|-R\safety$ and $b_{j''}$ was mined by an honest player. Let $r'_0$ denote the round when this block was mined.
    \item Note that at $r'_0, b_{j''}$ was mined by an honest player holding a chain of length $j'' \geq |C|-R\safety-\safety$; additionally, at $r$, $i$ is honest, holding a chain of length at most $|C|+\safety$ (recall that $|C|$ contains the blocks on which $i$ and $j$ agree, and by consistency, all but the last $\safety$ blocks in the chain of $i$ must be in the chain of $j$). Thus, by \emph{chain growth upper bound}, for any arbitrarily small $\delta'$ at most 
    $$\mu=(1+\delta')\frac{2\safety+R\safety}{np}$$
    rounds could thus have elapsed between $r'_0$ and $r$.
    \item By WLOG2, any \wo{} which gets added after $C$ must have been mined after $r'_0$. By WLOG4, any such \wo{} that is part of the chain of $j$ by $r'$ was mined before $r'$.
    \item We thus conclude by the Chernoff bound that except with a probability of $e^{-\Omega(np_f\cdot\frac{\safety(R+2)}{np})}=e^{-\Omega(q(R+2)\safety)}$, there were at most
    
    \begin{equation}    
    \label{eq:fruitgrowthupper}
    \begin{split}
        (1+\delta')^2\cdot{}np_f\cdot\Big(\frac{\safety(R+2)}{np}+t\Big)= \\ 
        (1+\delta')^2(q(R+2)\safety+np_ft)\leq\safetyfruit+(1+\delta')^2np_ft
    \end{split}
    \end{equation}

    ``new'' \wo{}s in the chain of $j$ at $r'$.
\end{itemize}

We conclude by noting that this implies a \wo{} growth upper bound of $g_1 = (1+\delta)np_f$ in the desired regime: Consider any $T\geq\frac{5\safetyfruit}{\delta}$ and any

$$t=\frac{T}{g_1}=\frac{T(1+\delta)}{np_f}$$

As shown above (see \cref{eq:fruitgrowthupper}), during this time $t$, except with negligible probability, the chain must have grown by at most 

$$\safetyfruit + (1+\delta')^2T(1-\delta)\leq T\delta/5+(1+\delta')^2 T/(1+\delta)$$

For any $0<\delta<1$ and $\delta' = 0.1\delta$, the above expression is upper bounded by $T$.
\fi

\subsection{\wo Fairness}

Disregard the chain growth, chain quality, \wo{} freshness, and \wo{} growth failure events, which happen with negligible probability\ifsubmission\else{} (see above)\fi. Consider  some $\phi$--fraction player subset selection $S$, some view $\view$ in the support of $EXEC^{\Pi_{\sys}(p,p_f,R)}(A, Z, \securityParam)$, some round $r$ and some player $i$ that is honest in round $r$ of $\view$. Let $C = b_0,\dots,b_{|C|}$ be the blocks in the view of $i$ at $r$, let $f_0,\dots,f_{\ell}$ be the \wo{}s contained in them, and let $m_0,\dots,m_{\ell}$ be the records contained in the \wo{}s; let $f_j,\dots,f_{j+T}$ be the $T$ consecutive \wo{}s for some $j$, where $T \geq \frac{5\safetyfruit}{\delta}$.

Let $r_0$ be the round when the \emph{block} in the view of $i$ at $r$ containing $f_{j+\safetyfruit}$ was first added to \emph{some} honest player $j_0$'s chain; let $r_1$ be the round when the block (again in the view of $i$ at $r$) containing $f_{j+T}$ was first added to some honest player $j_1$'s chain, and let $t = r_1 - r_0 - 2$ be the number of rounds from $r_0+1$ to $r_1-1$. We lower bound the number of $S$-compatible (honest) \wo{}s in the sequence, similar to the proof of \wo{} growth \emph{lower bound}:

\begin{itemize}
    \item By the \emph{\wo{} freshness} condition, every \wo{} mined by some honest player between $(r_0+1)$ and $(r_1-1)-wait$ will be in the chain of $j_1$ at some position $pos$ that is at least $\safety$ positions from the end of the chain, before the beginning of round $r_1$ and will remain so.
    \item By the Chernoff bound, in the $t-wait$ rounds from $r_0+1$ to $(r_1-1)-wait$, except with probability $e^{-\Omega((t-wait)\phi{}np_f)}$, the honest parties in $S$ mine at least $(1-\delta')(t-wait)\phi{}np_f$
    \wo{}s (where $\delta'$ is some arbitrarily small constant), which thus are all included in the chain of $j_1$ by $r_1-1$.
    \item Since \wo{}s are ordered by the block containing them, and since in round $r_1$ a \emph{new} block is added which contains $f_{j+T}$, it follows from \emph{blockchain consistency} that all these \wo{}s contained in the sequence $f_1,\dots,f_{j+T}$ (recall that all these \wo{}s are found in blocks that are at least $\safety$ positions from the end of the chain, so by consistency, those blocks cannot change and thus were not added in round $r_1$ and consequently must come before the block containing $f_{j+T}$).
    \item \ifsubmission\else By WLOG3 (cf. \cref{sec:simplif-assump})\fi, these \wo{}s are also all ``new'' (i.e., not included in the chains of $j_0$ at $r_0$) and different. Since in round $r_0$, the block containing $f_{j+\safetyfruit}$ was added to the chain of $j_0$, and \ifsubmission\else{}since by WLOG5\fi, the chain of $j_0$ at $r_0$ up until (and including) the block which contains $f_{j+\safetyfruit}$ is a prefix of $C$, all these \wo{}s must in fact be contained in the sequence $f_{j+\safetyfruit},\dots,f_{j+T}$.
    \item Finally, by \wo{} consistency, at $r_0$ all honest players' \wo{} chains contain $f_1,\dots,f_j$ (since recall that some player added $f_{j+\safetyfruit}$ at $r_0$). Thus, all these \wo{}s are $S$-compatible w.r.t. the prefix $f_1,\dots,f_{j-1}$ before the $T$ segment we are considering.
\end{itemize}

We proceed to show that $t$ is sufficiently large. Recall that $j_0$ is honest at $r_0$ and $j_1$ is honest at $r_1$. We know that at $r_1$, the \wo{} chain contains at least $f_{j+T}$ \wo{}s. Additionally, at $r_0$ the \wo{} $f_{j+\safetyfruit}$ is added for the first time, so by \wo{} consistency, at most $j+2\safetyfruit$ \wo{}s could have been in the chain of $i$ at this point (since a \wo{} at position $j+\safetyfruit$ is modified). Thus, the \wo{} chain must have grown by at least $T-2\safetyfruit$ from $r_0$ to $r_1$. By the \emph{upper bound on \wo{} growth} we have that: $T-2\safetyfruit\leq\safetyfruit+(1-\delta')^2np_f(t+2)$.
Thus: $t\geq\frac{1}{(1+\delta')^2np_f}(T-3\safetyfruit)-2$.

We conclude that (except with negligible probability) the number of \wo{}s in the sequence is at least:
\begin{align*}
(1-\delta')\phi{}np_f\Big(\frac{1}{(1+\delta')^2np_f}(T-3\safetyfruit)-2-wait\Big)= \\
(1-\delta')\phi\Big(\frac{1}{(1+\delta')^2}(T-3\safetyfruit)-np_f(wait+2)\Big)\geq \iffull \\ \else\fi
(1-\delta')\phi\Big(\frac{1}{(1+\delta')^2}(T-3\safetyfruit)-\safetyfruit\Big)\geq \\
(1-\delta')\phi\Big(\frac{1}{(1+\delta')^2}(T-4.5\safetyfruit)\Big)\geq 
\phi(T-5\safetyfruit)
\end{align*}
\ifsubmission
\else
where the first inequality follows from Fact \ref{fact1}, and the second and third by taking a sufficiently small $\delta'$. 
\fi
Since $T\geq\frac{5\safetyfruit}{\delta}$, we have that $T-5\safetyfruit \geq (1-\delta)T$, thus the number of \wo{}s in the sequence is at least: $(1-\delta)\phi{}T$.

\subsection{Incentive Compatibility}

Our final theoretical result is game-theoretic and concerns utility-maximizing participants.

Any secure blockchain protocol that satisfies $\delta$-approximate fairness (where $\delta < 0.3$) w.r.t $T(\kappa)$ length windows can be used as the ledger underlying a cryptocurrency system
while ensuring $3\delta$-incentive compatibility if players (i.e., miners) only care about how much money
they receive.

We say that honest mining is a $\rho$-coalition-safe $\epsilon$-Nash equilibrium if, with overwhelming probability, no $\rho' < \rho$ fraction coalition can gain more than a multiplicative factor $(1 + \epsilon)$ in utility,
no matter what transactions are being processed; formally, consider some environment providing transactions into the system. We restrict to a setting where the total rewards and transaction fees during the run of the system are some fixed constant $V$.

Fairness implies that no matter what deviation the coalition performs, with overwhelming probability, the fraction of adversarial blocks in any $T(\kappa)$-length window of the chain is upper bounded by $(1+\delta)\rho$ and thus the total amount of compensation
received by the attacker is bounded by $(1 + \delta)\rho \cdot V$ ; in contrast, by fairness, if the coalition had
been following the honest protocol, they are guaranteed to receive at least $(1 - \delta)\rho \cdot V$; thus, the
multiplicative increase in utility is:
$\frac{1 + \delta}{1 - \delta} \leq 1 + 3\delta$
when $\delta < 0.3$.

\section{Evaluation}\label{sec:evaluation}

We now compare the Proportional Reward Splitting (PRS) mechanism to other
reward mechanisms under realistic parameters. In order to perform the
comparison, we utilized the Markov Decision Process (MDP) implementation
of~\cite{DBLP:conf/sp/0003P19}. MDPs are fitting tools in our case since we
want to evaluate the optimal strategies \wrt a specific utility. Therefore,
when parameterizing our MDPs with realistic values, \eg in terms of block
finality, we can identify the optimal strategy that utility-maximizing rational
parties would opt for.

The model does not require knowledge of the identities of parties, or which
parties are adversarial, but only assumes that the adversary $\adversary$ holds a certain
amount of power. In practice, a protocol's designer could use our analysis to
get a lower bound on their protocol's performance, by pinpointing the advantage
of the strongest adversary that their protocol can tolerate.

Next, we first outline the construction of the MDP
and then compare PRS with other reward mechanisms under three metrics, namely
incentive compatibility, subversion gain, and censorship susceptibility.
We then evaluate how different values of the PRS's parameters, namely the
workshare and fork eligibility windows, affect its performance. Finally, we
analyze the accuracy of estimating the adversarial and honest parties' power
via workshares and outline the tradeoff between sampling accuracy and storage
overhead.

\subsection{Description of Markov Decision Process}

The MDP implementation of PRS is a direct adaptation of the vanilla Reward
Splitting (RS) MDP. Here we give a brief description of the MDP and refer to
\cite{DBLP:conf/sp/0003P19} for more details. Our implementation is available at~\cite{evalrepo}.

We note that our evaluation is not a Monte Carlo simulation with independent runs. For each
parameter point, we construct the corresponding finite MDP and solve for the
optimal adversarial policy under the selected reward function. Therefore, the reported
values correspond to expected long-run rewards induced by the optimal policy,
rather than empirical means over repeated executions.

The state of the MDP is a tuple of four elements $\langle l_a, l_c,
\text{fork}, \text{history} \rangle$: 
\begin{itemize}
    \item $l_a$ denotes the length of the (private) adversarial chain;
    \item $l_c$ denotes the length of the honest (public) chain;
    \item $\text{fork}$ takes three values: \emph{active} if
        there is a tie, \ie honest miners are split between two
        chains; \emph{cLast} if the latest block is honest; \emph{aLast} if the
        latest block is adversarial.
    \item $\text{history}$ is a bitstring as follows: (i) its length represents
        the number of consecutive attacker blocks in the common (main) chain;
        (ii) each position corresponds to a height of the main chain, starting
        from the last block (the tip of the chain); (iii) a $0$ bit denotes
        that no competing honest block exists for the corresponding height,
        whereas $1$ means that an honest block for that height exists.
\end{itemize}

State transitions occur via the following actions:
\begin{itemize}
    \item \emph{Adopt}: $\adversary$ adopts the public chain.
    \item \emph{Wait}: $\adversary$ keeps mining privately and publishes no blocks.
    \item \emph{Match}: $\adversary$ publishes a chain of private blocks equal
        to the length of the public chain, causing a tie; this is feasible when
        $l_a \geq l_c$ and $\text{fork} = \text{cLast}$.
    \item \emph{Override$_k$}: $\adversary$ publishes a chain of private blocks
        which is $k$ blocks longer than the public chain.
\end{itemize}

Finally, the reward allocation is the point of difference between PRS and RS.
In RS, when choosing \emph{Adopt} the honest miners and $\adversary$ each
receive half of the rewards for heights where both honest and adversarial
blocks exist. Instead, in PRS the rewards are split proportionally
to each side's mining power; for instance, if the honest miners (are set to)
have $70$\% of mining power, they receive $70$\% of the rewards for these
heights, instead of $50$\% as in RS.\footnote{In practice it is impossible to
know the exact power distribution between honest parties and the adversary.
Instead, an approximation is made using workshares. The accuracy of this
approximation is explored in Section~\ref{sec:approximation}.} For the heights
where only honest blocks exist, the honest miners get the full reward.
Additionally, the adversary receives the full reward for heights where honest
objects (blocks and/or workshares) are ``pushed out'' of the history, \ie the
adversary creates enough consecutive blocks to censor their inclusion.

\subsection{Comparison of Reward Mechanisms}

We first explore the metrics established in~\cite{DBLP:conf/sp/0003P19},
namely:
(i) incentive compatibility;
(ii) subversion gain;
(iii) censorship susceptibility.
For each metric, we compare PRS with Bitcoin, FruitChains, and vanilla RS.

In all evaluations, we used the same parameters for the eligibility windows.
The first window concerns object eligibility (i.e., the recency window), that is the number of block heights
within which an object should be published in order to be eligible for rewards.
This window was set to $\shareEligibilityWindow=6$. The second window concerns
fork eligibility, that is the depth of a fork up to which objects (that point
to blocks of that depth) are eligible for rewards. This window was set to
$\forkEligibilityWindow=6$.

The parameter $\selfishMiningNetworkParam$, which indicates the percentage of
honest miners that adopt the adversarial forks, is set to $0.5$ for Bitcoin,
RS, and PRS, and $0$ and $1$ for
FruitChains.\footnote{For FruitChains we use both $0$ and $1$ because the
FruitChains MDP implementation of~\cite{DBLP:conf/sp/0003P19} did not enable
using $0.5$ due to complexity issues.}
In essence, when two chains of the same height are available, $\selfishMiningNetworkParam = 0.5$ implies that honest miners choose between
competing chains of equal height uniformly at random.

Following the baseline configuration of~\cite{DBLP:conf/sp/0003P19}, we set the fruit-to-block ratio to 1. While higher fruit rates improve the incentive compatibility of FruitChains, at the cost of increased bandwidth,~\cite{DBLP:conf/sp/0003P19} observes only small improvements for higher ratios and finds RS to outperform FruitChains in its comparison.

\mypar{Incentive Compatibility}
\begin{figure}
    \begin{center}
        \includegraphics[width=\iffull{}\fi\columnwidth]{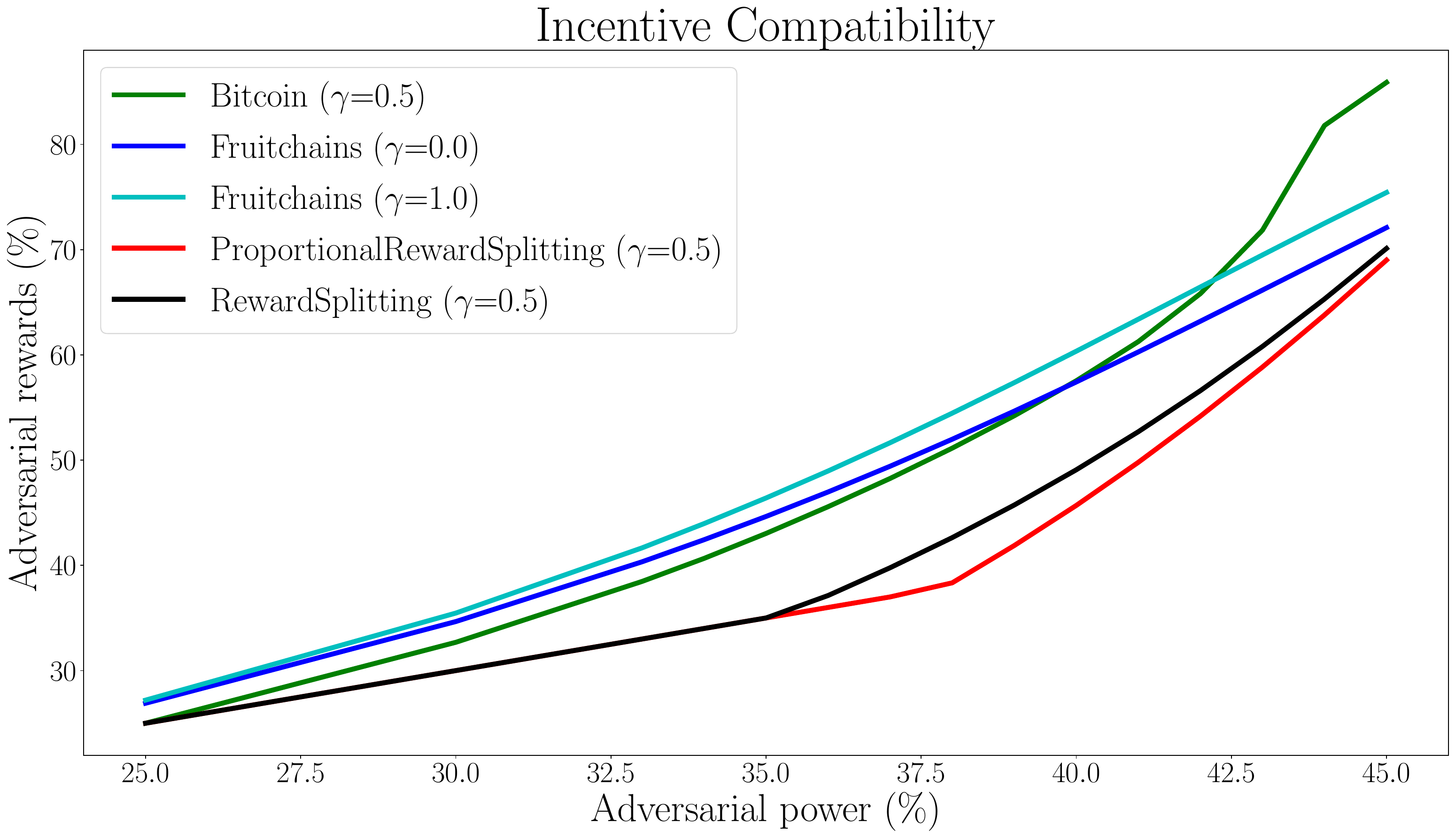}
    \end{center}
    \caption{
        Incentive compatibility for various reward mechanisms.
        The eligibility window for fruits (in FruitChains) and objects (in
        Proportional Reward Splitting) is set to $\shareEligibilityWindow=6$.
        The fork eligibility window for FruitChains, Reward Splitting, and
        Proportional Reward Splitting is set to $\forkEligibilityWindow=6$.
        Larger values indicate worse performance.
    }
    \label{fig:ic}
\end{figure}
The first metric of comparison is incentive compatibility (IC). Briefly, IC
expresses the percentage of rewards that the adversary gains compared to its
fair share, that is, its relative mining power. \cref{fig:ic} depicts the IC for
the reward mechanisms under question. In essence, values above the line $x=y$
indicate that the adversary gains disproportionately more rewards, compared to
its mining power, so larger values indicate worse IC performance.

PRS outperforms all other mechanisms on IC. For adversarial power up to $25\%$,
all mechanisms except FruitChains are optimal. RS and PRS both remain optimal
through $35\%$; beyond that, PRS strictly outperforms RS.

\mypar{Subversion Gain}
\begin{figure}
    \begin{center}
        \includegraphics[width=\iffull{}\fi\columnwidth]{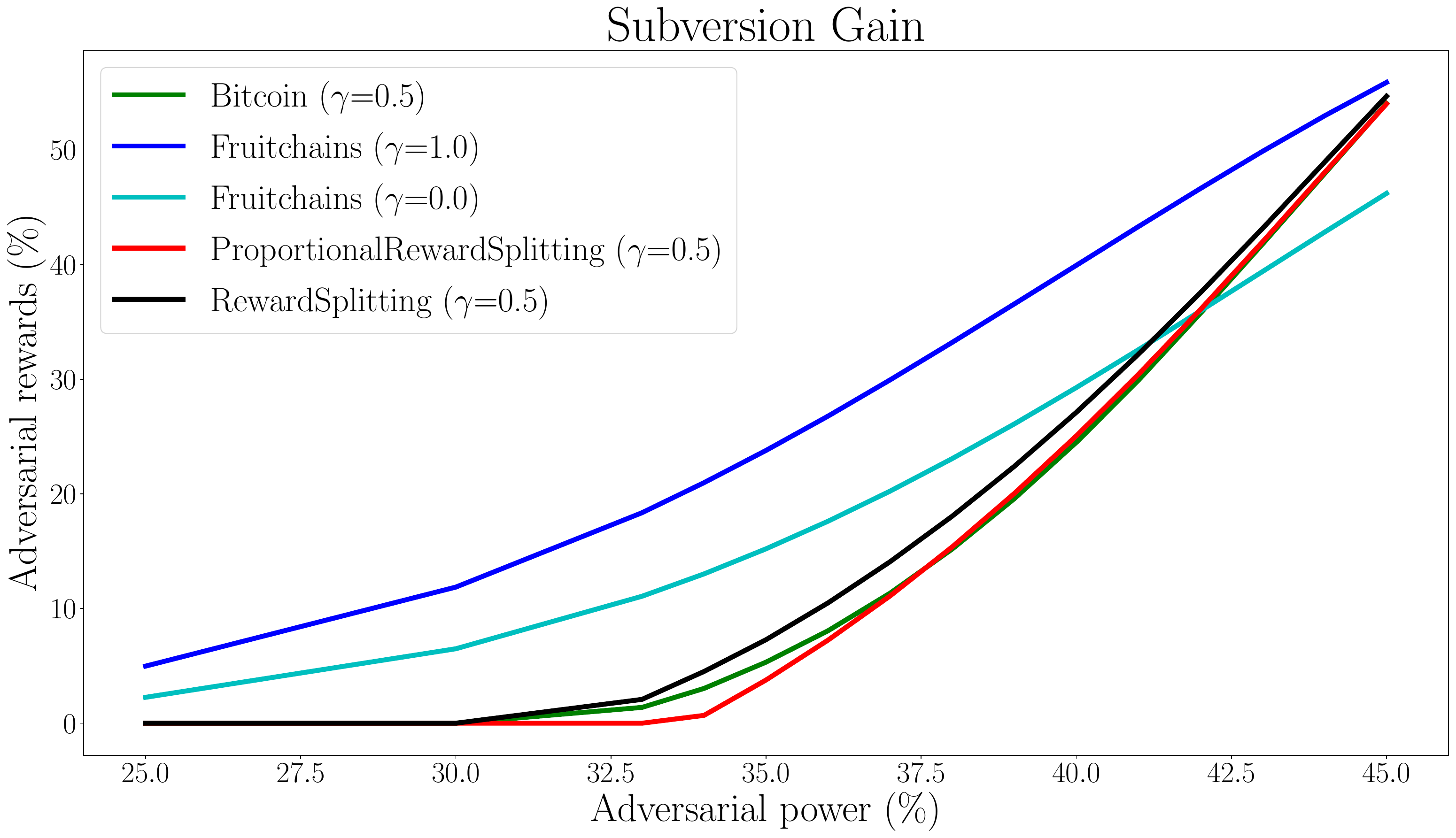}
    \end{center}
    \caption{
        Comparison of subversion gain for various reward mechanisms.
        The eligibility window for fruits (in FruitChains) and objects (in
        Proportional Reward Splitting) is set to $\shareEligibilityWindow=6$. 
        The fork eligibility window for FruitChains, Reward Splitting, and
        Proportional Reward Splitting is set to $\forkEligibilityWindow=6$.
        Larger values indicate worse performance.
    }
    \label{fig:subversion}
\end{figure}
Subversion gain indicates the profitability of double spending attacks. In
particular, the profit from a double-spending attack is time-averaged and set
to $3$ times the block reward. In other words, if the adversary manages to
orphan $k$ blocks during the attack, it is granted $3 \cdot k$ rewards.  The
confirmation time is set to $6$ blocks, so the attack is successful if the
adversary reverts $6$ blocks in order to successfully double-spend the assets.
If the attack is unsuccessful, the adversary gets no rewards, but also incurs
no penalty.

\cref{fig:subversion} depicts the comparison of the reward mechanisms \wrt
subversion gain. For the most part, PRS is again the
better option. In particular, up to $30$\% all mechanisms except FruitChains
perform optimally, \ie the adversary has zero gain. For the range of
adversarial power between $30$\% and $38$\% PRS is
the best option, whereas between $38$\% and $42$\% it is in effect equally good
to Bitcoin. Above $42$\%, FruitChains with $\selfishMiningNetworkParam=0$
performs better, which is a result of the particularly favorable setting
(compared to $\selfishMiningNetworkParam=0.5$ for the other mechanisms);
instead, under $\selfishMiningNetworkParam=1$, FruitChains is strictly worse
than all others for all adversarial powers.

In summary, PRS is optimal \wrt subversion gain and,
in some cases, it is strictly better than all other options.

\mypar{Censorship Susceptibility}
\begin{figure}
    \begin{center}
        \includegraphics[width=\iffull{}\fi\columnwidth]{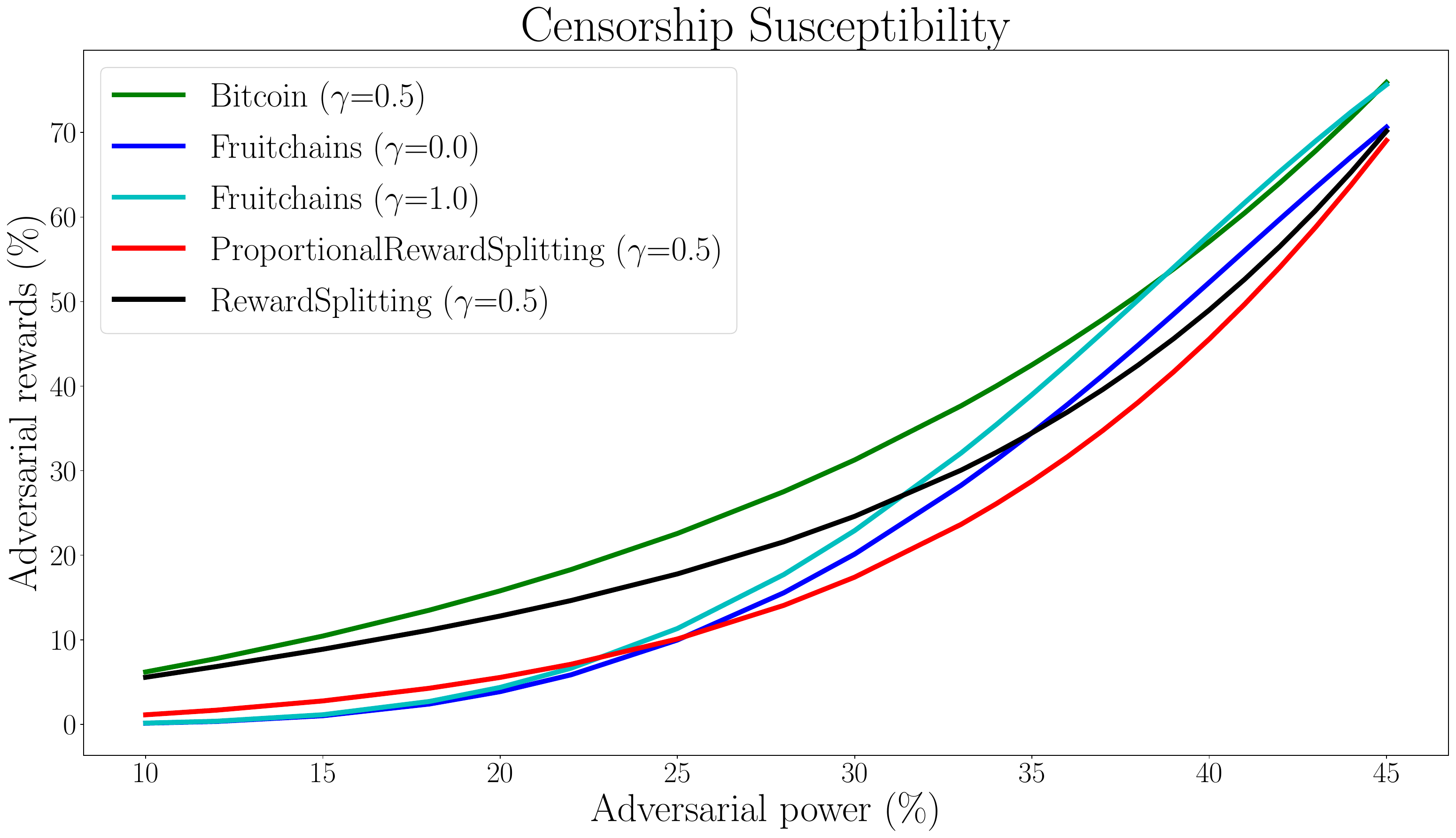}
    \end{center}
    \caption{
        Comparison of censorship susceptibility for various reward mechanisms.
        The eligibility window for fruits (in FruitChains) and objects (in
        Proportional Reward Splitting) is set to $\shareEligibilityWindow=6$. 
        The fork eligibility window for FruitChains, Reward Splitting, and
        Proportional Reward Splitting is set to $\forkEligibilityWindow=6$.
        Larger values indicate worse performance.
    }
    \label{fig:censorship}
\end{figure}
Censorship susceptibility indicates the maximum fraction of income loss that
the adversary can incur on honest miners under the threat of a censorship
attack. Specifically, the adversary's reward corresponds to the honest miners'
loss in the form of orphaned blocks.

\cref{fig:censorship} depicts the results for censorship susceptibility. These
are consistent with~\cite{DBLP:conf/sp/0003P19}, wherein FruitChains
outperforms the other mechanisms for smaller adversaries but
becomes relatively worse for larger values. Interestingly,
PRS strictly outperforms vanilla RS for all adversarial
powers. As a result, FruitChains is the best option only for adversarial power
up to $25$\%, after which point PRS is the best
option. Notably, even for smaller values of adversarial power, the difference
between PRS and FruitChains is less than $2$\%.

\subsection{Comparison of Eligibility Parameters}

We now evaluate how the different eligibility parameters affect the performance
of the reward mechanisms. In particular, we will compare the two types of
reward splitting \wrt the fork and the workshare eligibility windows. The
metric of comparison is incentive compatibility, since all metrics are affected
in the same manner when changing the parameters. As before, we set
$\selfishMiningNetworkParam$ to be $0.5$.

\mypar{Workshare Eligibility Window}
The first parameter is the workshare eligibility window
$\shareEligibilityWindow$. This is the distance between the
block that a workshare references and the block within which it gets published.
Intuitively, larger window values require the adversary to censor the chain for
longer periods of time, in order to force an honest workshare to become
ineligible, thus making the attack harder. However, they also incur a delay on
the finalization of objects and, consequently, the allocation of rewards.

\begin{figure}
    \begin{center}
        \includegraphics[width=\iffull{}\fi\columnwidth]{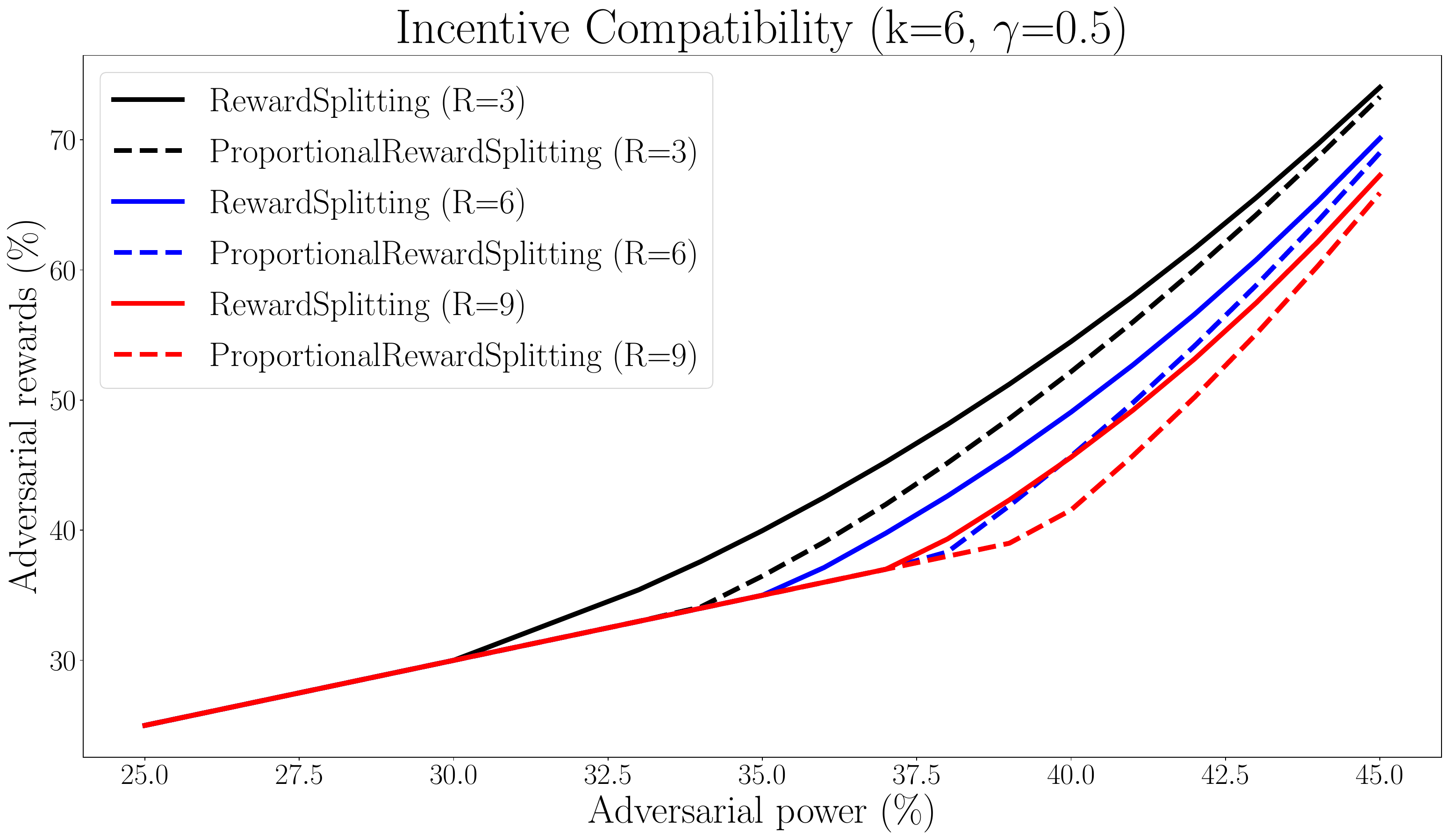}
    \end{center}
    \caption{
        Incentive compatibility \wrt the
        workshare eligibility window between Reward Splitting and Proportional
        Reward Splitting. The fork window for objects is set to $\forkEligibilityWindow=6$. 
        Larger values indicate worse performance.
    }
    \label{fig:ic-comparison-k}
\end{figure}

\cref{fig:ic-comparison-k} depicts how the two reward splitting mechanisms
perform for three workshare eligibility window values, $\shareEligibilityWindow
\in \{3, 6, 9\}$. Evidently, Proportional Reward Splitting outperforms vanilla
Reward Splitting in all cases. As expected, larger values of
$\shareEligibilityWindow$ result in better performance, that is, lower
adversarial rewards. Consequently, PRS is optimal for larger values of
adversarial power; when $\shareEligibilityWindow = 3$, PRS is optimal for
adversarial power up to $33$\%, whereas for $\shareEligibilityWindow = 9$ this
goes up to $38$\%.

\mypar{Fork Eligibility Window}
The second parameter is the fork eligibility window $\forkEligibilityWindow$.
This window sets the depth of a fork up to which objects are eligible for
rewards. Specifically, if $\forkEligibilityWindow = 3$, then objects that
point to a block in a fork are eligible for rewards only if this block is among
the first $3$ of the fork.
Note that the height of the workshare, which is considered for the workshare
eligibility window, is computed as the height of the referenced block. For
example, if a workshare points to the second block of a fork which occurred at
height $100$, then the workshare's height is $102$ and, in order to be eligible
for rewards, it should be published in a (main chain) block up to height $102 +
\shareEligibilityWindow$.

\begin{figure}
    \begin{center}
        \includegraphics[width=\iffull{}\fi\columnwidth]{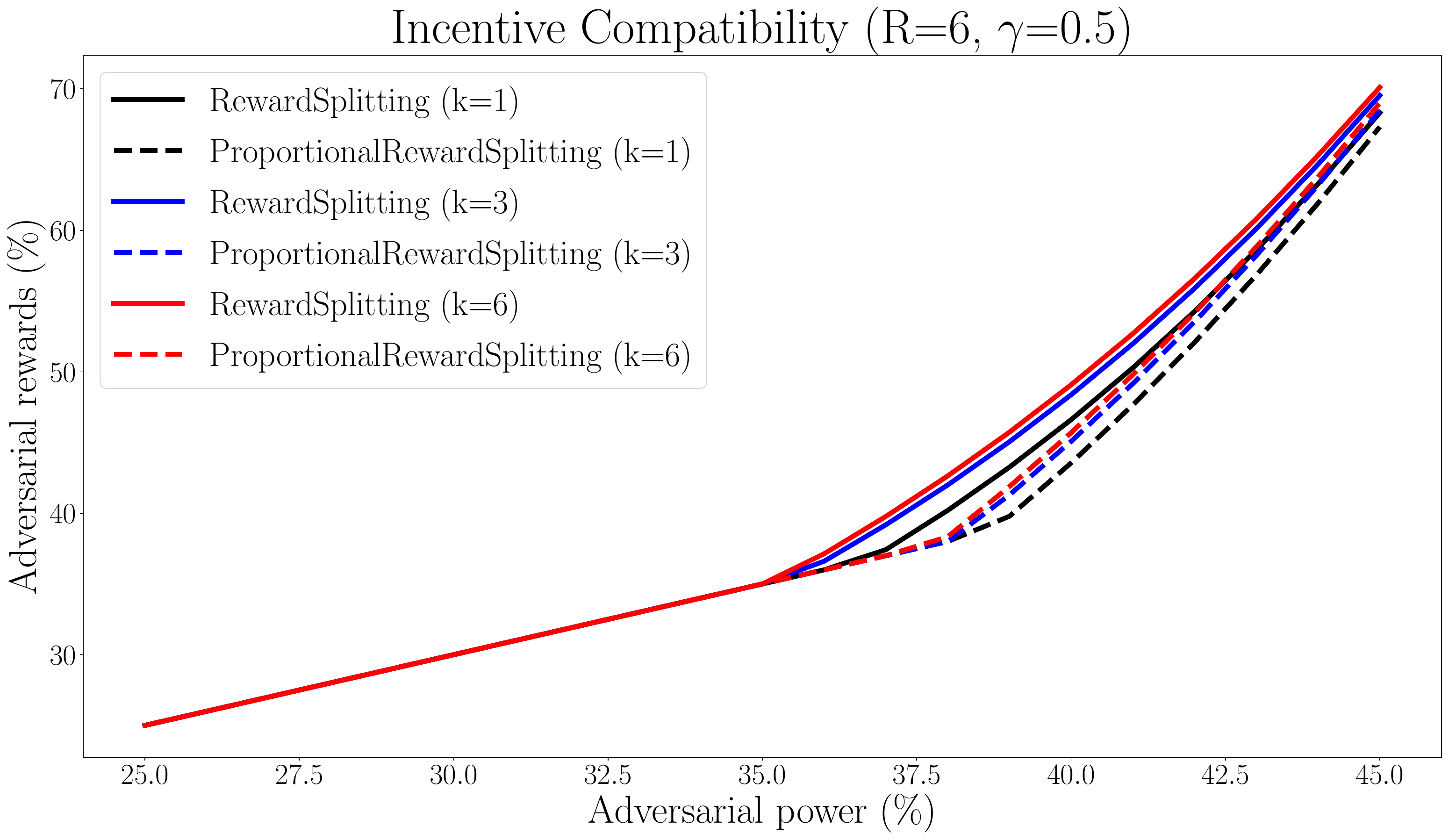}
    \end{center}
    \caption{
        Incentive compatibility for various values of the fork
        window between Reward Splitting and Proportional Reward
        Splitting. The eligibility window for objects is set to $\shareEligibilityWindow=6$. 
        Larger values indicate worse performance.
    }
    \label{fig:ic-comparison-w}
\end{figure}

\cref{fig:ic-comparison-w} depicts the effect of the fork eligibility window on
incentive compatibility. Evidently, lower values of $\forkEligibilityWindow$
result in better performance, \ie lower adversarial rewards. One explanation
for this could be that lower values of $\forkEligibilityWindow$ penalize
forking more, so the adversary is incentivized to abandon forks more often,
hence its rewards get closer to its fair share.

Interestingly, we observe that PRS outperforms RS even for different values of
$\forkEligibilityWindow$. Specifically, PRS under $\forkEligibilityWindow=6$
performs better (for the most part) than RS under $\forkEligibilityWindow=1$.

\subsection{Sampling Accuracy}\label{sec:approximation}


So far, our evaluation of the proportional reward splitting mechanism assumes a
perfect knowledge of the distribution of power between the honest parties and
the adversary.
In practice, this is not possible to achieve for proof-of-work (PoW)
ledgers, as each party's proportional power is not visible to
anyone, not even to the party itself. Although a party knows its
\emph{absolute} power, they do not know the exact amount of power of other
parties.

Instead, the distribution of power can be inferred from observed events,
specifically block creations. Since the probability that each party creates a
block on each round is proportional to their power and each such event is
independent, a large enough sample of blocks enables inferring the power
distribution.

A pertinent question is how many samples are needed to make an accurate enough
estimation, with a low margin of error. In our setting, the samples
correspond to workshares and blocks. Therefore, the question of interest is how
many workshares should be published per height, such that each block's rewards are
distributed proportionally to each side's power, accurately and in most cases
(low margin of error).

To estimate the number of workshares needed, we model the generation of workshares as a
series of independent Bernoulli trials. Let $X_i \in \{ 0, 1 \}$ be the i-th
sample of random variable $X$, where $0$ corresponds to an adversarial workshare
creation and $1$ corresponds to honest workshare. The probability of $1$ for each
$X_i$ is $p_h$, so $\{ X_i \}_{i \in [n]}$ is independent and identically
distributed to a Bernoulli trial $\textsf{Bern}(p_h)$.

Let $X = \sum_{i=1}^{n} X_i$ be the random variable that represents the sum of
the observed values, \ie the number of honest workshares. It holds that $E[X] = n
\cdot p_h$, where $n$ is the number of samples (\ie Bernoulli trials) and $p_h$ is
the probability of creating an honest workshare.

By the Chernoff bound:
$\text{Pr}\left[ X \leq (1-\delta) \cdot E[X] \right] < \epsilon = e^{-\frac{\delta^2 \cdot n \cdot p_h}{2}}$.
where:
(i) $\delta$ denotes the acceptable deviation from the completely accurate
(expected) value; (ii) $\epsilon$ denotes the error bound of the observation,
\ie the probability that the observed value is less accurate than $\delta$.
By solving for $n$ we find, for a given honest-power fraction $p_h$, the
minimum number of workshares needed to get an accurate enough observation with
low error probability:
$n = - \frac{2 \cdot \log \epsilon}{\delta^2 \cdot p_h}$.

\begin{figure}
    \begin{center}
        \includegraphics[width=\iffull{}\fi\columnwidth]{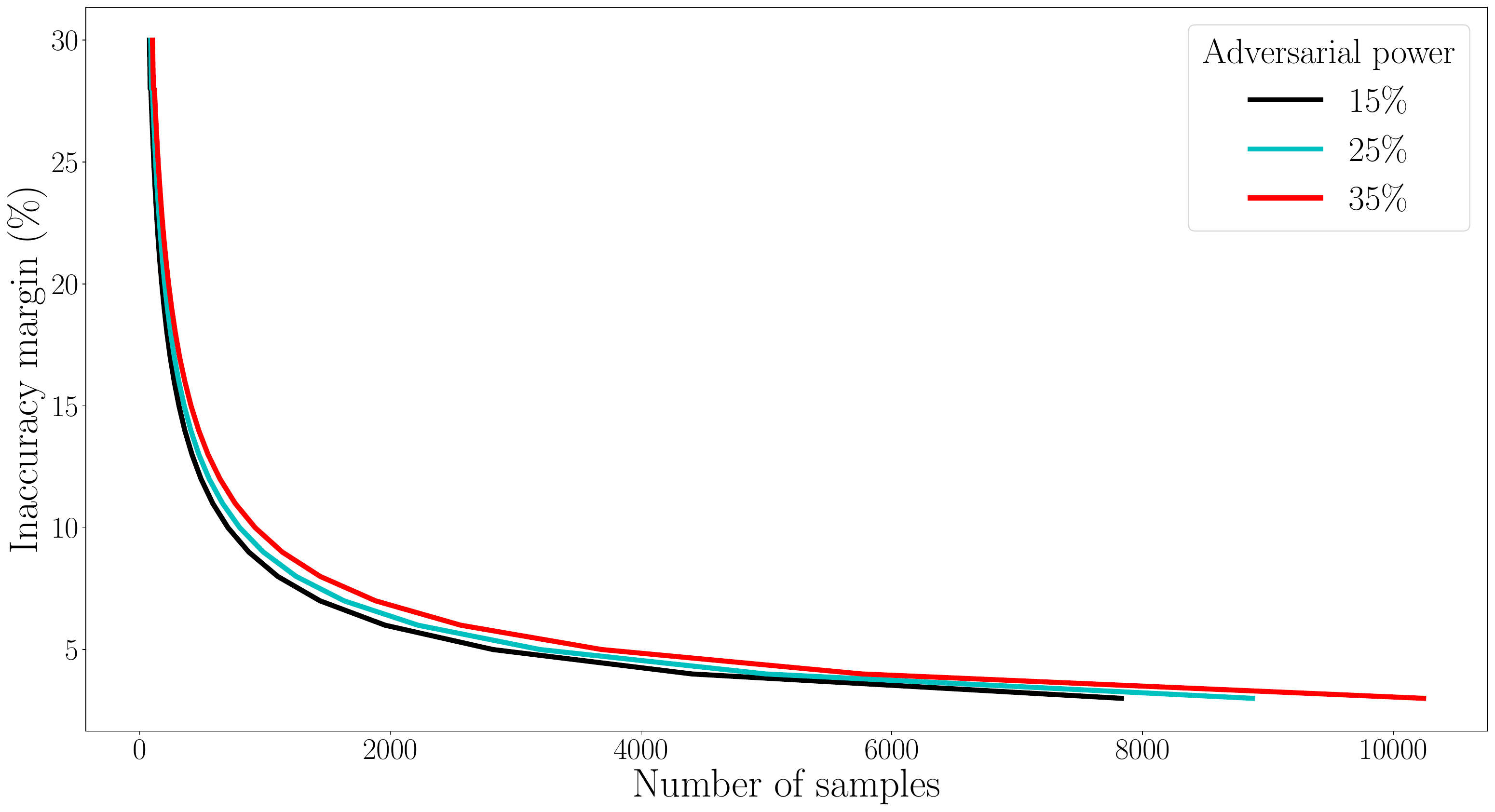}
    \end{center}
    \caption{
Number of workshares needed to estimate the honest/adversarial power distribution, at error bound $\epsilon = 0.05$. Inaccuracy is the acceptable deviation from the ideal value. 
    }
    \label{fig:samples}
\end{figure}

\cref{fig:samples} depicts how $n$ is affected given various values of the
other parameters. Here, we set the error bound to $0.05$; in other words, the
probability that the observed distribution between the honest and adversarial
parties is less accurate than the acceptable level of inaccuracy is $5$\%.

The x-axis corresponds to the number of workshares, while the 
y-axis
corresponds to the acceptable level of inaccuracy. We observe that inaccuracy
drops exponentially \wrt the number of samples used. Additionally, inaccuracy
drops at a faster rate when the adversarial power is lower. 
We assume that each workshare is equivalent to a Bitcoin block's
header, which amounts to $80$ bytes. Therefore, the overhead \emph{per block}
increases linearly with the number of workshares. As we explain in \cref{sec:removestorage}, however, this storage is off-chain and can be discarded after finalizing the rewards. 

For example, assume adversarial power is $15$\%. $7{,}832$ workshares achieve $3$\%
inaccuracy (with error probability $5$\%), requiring $612$ KB
per block. Similarly, $1{,}000$ workshares achieve approx. $8.5$\% inaccuracy
at $79$ KB. On the other hand, assuming $35$\% adversarial power, $3$\%
inaccuracy is achieved with $10{,}242$ workshares, which corresponds to $800$ KB.

\section{Discussion and Deployment}
\label{sec:discussion}


\mypar{PRS in theory} PRS matches FruitChains in theoretical guarantees but outperforms it in
practical settings. This is because the theoretical analysis
evaluates the protocols \emph{asymptotically}, for (very) large
parameters. In practice, parameters like freshness or the fork
eligibility window cannot be too large without delaying transaction and reward
finalization.

At realistic parameter values, the theoretical guarantees no longer apply, so
we turn to the MDP evaluation. One explanation for why PRS outperforms
FruitChains lies in the tradeoff between ``Rewarding the Bad'' and
``Punishing the Good'' identified in~\cite{DBLP:conf/sp/0003P19}.
FruitChains eliminates the incentive to fork by rewarding the adversary for
failed attempts. PRS does the same through workshares, but to a lesser degree:
it splits rewards proportionally rather than yielding the full contested reward
to the adversary. This finer split strikes a better balance in practice.

\mypar{PRS in practice} PRS outperforms Reward Splitting (RS) in practical terms.
Both mechanisms follow the same punishment strategy, that is, 
splitting the rewards, but PRS does so in a more fine-grained way, which better approximates the distribution of power between the adversary and the honest parties.
In other words, the gain of PRS comes from the difference between $50$\% and
the actual value of adversarial power; hence, when the latter approaches the
former, the two lines converge.


\mypar{Dynamic difficulty}
Our theoretical analysis follows the methodology \iffull{}of~\cite{DBLP:conf/podc/PassS17}\fi{}, which operates in the
static-difficulty setting; we therefore do not establish formal guarantees for
variable-difficulty executions. We expect the same intuition to apply in this
setting, similarly to variable-difficulty analyses of accumulated-work
protocols such as NIPoPoWs: when the target changes, work should be interpreted
relative to the target that was active when the object was created. In a
deployment, the workshare target would be recalculated using the same
difficulty-adjustment rule as the main block target, so that workshares
continue to provide a fresh sample of the mining-power distribution. A formal
analysis of PRS under variable difficulty is left for future work.

\mypar{Deployment of \protocol} Our \protocol transformation has been deployed in practice as part of a Proof-of-Work blockchain launched in January 2025~\cite{quaiwsrepo} that uses PoEM as its fork-choice rule. The live blockchain records workshares on-chain with a hard cap of 9 workshares per block and a soft target of 3, where each workshare header is approximately 120 bytes, which is comparable to a standard transaction header. Workshares older than 2 blocks (freshness parameter) are rejected by peers. Rewards are paid proportionally to each miner's workshare-estimated power, with a linear staleness discount applied to shares approaching the freshness timeout, incentivizing timely propagation. At the current parameters, the storage overhead amounts to under 1 KB per block above baseline.

\ifsubmission\else
\fi
\section{Related Work}

The work most directly related to \protocol is
FruitChains~\cite{DBLP:conf/podc/PassS17}.
FruitChains was shown to be a
$\rho$-coalition-safe $\epsilon$-Nash equilibrium, \ie any coalition of parties
with a fraction of power less than $\rho$ cannot gain more than $\epsilon$ of
their fair share of rewards. As shown in \cref{sec:analysis}, our
mechanism achieves the same guarantee, as the analysis of FruitChains carries
over in our case almost directly. Nonetheless, \protocol outperforms FruitChains in
practice, that is, for parameter values smaller than the theoretical,
unrealistic, bounds (cf. \cref{sec:evaluation}).

\iffull{}The recording of \fi{} low-work objects on-chain was explored in
StrongChain~\cite{DBLP:conf/uss/SzalachowskiRHS19}. StrongChain
publishes workshares on-chain, as opposed to keeping them within
a mining pool and invisible to the protocol. The paper offers a
protocol implementation and a brief discussion of the
protocol's performance in the optimistic setting, where all parties are honest.
However, StrongChain's analysis is lacking in many aspects.
Specifically, StrongChain offers only a high-level description of reward allocation
and an intuition of why this allocation is beneficial. In
comparison, our mechanism is analyzed both in theory, in a game-theoretic
setting where parties are assumed rational instead of simply honest, and in
practice via Markov decision processes. Moreover, StrongChain's
analysis considers only a specific instance of a selfish mining strategy, and
analyzes the protocol's performance under various parameter values, whereas our
MDP analysis evaluates the performance of the optimal adversarial strategy.

Another line of work on recording low-work objects on-chain
builds on the idea of ``weak blocks''. These works include
Subchains~\cite{DBLP:journals/ledger/Rizun16} and
Flux~\cite{DBLP:journals/iacr/ZamyatinSSWK18}. Here, the main idea is the
formation of ``sub-chains'', that is, chains of objects with less PoW
than blocks. Therefore, between two consecutive (heavy) blocks, a sub-chain of
weak blocks may be created, enabling a more fine-grained estimation of the
power distribution among participants. Flux offered a simulation-based
analysis, showing that it performs better \wrt selfish mining compared to
Bitcoin. However, as was shown in~\cite{DBLP:conf/sp/0003P19}, Subchains
performs worse in practice compared to Reward Splitting (RS). In comparison,
our mechanism outperforms RS, and thus also Subchains, under the framework
of~\cite{DBLP:conf/sp/0003P19}.

The uncle reward system used in Ethereum~\cite{buterin2014ethereum} prior to its transition to Proof-of-Stake rewarded non-canonical work via \emph{uncle blocks}. Ritz and Zugenmaier~\cite{RZ2018} show empirically that such a fixed partial reward is insufficient: it actually \emph{lowers} the mining power threshold at which selfish mining becomes profitable, compared to Bitcoin, rather than eliminating the incentive. The core issue is that a fixed reward subsidizes the adversary regardless of their actual mining power. \protocol avoids this by splitting rewards proportionally to the power distribution, ensuring the adversary cannot benefit beyond their fair share.

Finally, \protocol can be instantiated on any longest-chain protocol that guarantees
safety and liveness, including Nakamoto Consensus~\cite{nakamoto2008bitcoin},
uniform tie-breaking~\cite{DBLP:conf/nsdi/EyalGSR16,DBLP:journals/cacm/EyalS18},
smallest-hash tie-breaking~\cite{lerner2015decor}, unpredictable deterministic
tie-breaking~\cite{DBLP:conf/uss/Kokoris-KogiasJ16},
Publish-or-Perish~\cite{DBLP:conf/ctrsa/ZhangP17}, and
PoEM~\cite{poem}. The security analysis requires only that the host protocol
satisfies these two properties; no other assumptions on the fork-choice rule
are needed.



\iflipics
    \bibliographystyle{plainurl}
\else
    \bibliographystyle{format/splncs04}
\fi
\bibliography{pubs}

\ifsubmission\else

\iffull

\appendix
\section{Chain Selection and Ledger Extraction}
Here, we present two algorithms omitted from the main body: the
ledger-extraction function (Algorithm~\ref{alg.extract}), 
and an example instantiation of $\maxvalid$ (Algorithm~\ref{alg.poem-maxvalid}).
We illustrate $\maxvalid$ with PoEM~\cite{poem}, which selects the chain with
the most accumulated intrinsic work. PoEM is the natural pairing because it
uses the same intrinsic-work measure that $\protocol$ uses for reward
weighting; any other host protocol's chain-selection rule applies as well,
including Bitcoin's longest-chain rule.  

\begin{algorithm}[ht]
    \caption{\label{alg.extract} The ledger extraction mechanism}
    \begin{algorithmic}[1]
        \Function{$\extract_{\safetyParam, \genesis}$}{$\chain$}
            \If{$\lnot \validate_{\genesis}(\chain)$}
                \State \Return $\bot$
            \EndIf
            \Let{\ledger}{[]} \Comment{Initialize empty ledger}
            \For{$\block \in \chain[:-\safetyParam]$} \Comment{Parse each stable block}
                \For{$\tx \in \block.\txList$}
                    \Let{\ledger}{\ledger \concat \tx.\payload}
                \EndFor
            \EndFor
            \State \Return{$\ledger$}
        \EndFunction
    \end{algorithmic}
\end{algorithm}
\begin{algorithm}[ht]
    \caption{\label{alg.poem-maxvalid} The PoEM maxvalid algorithm}
    \begin{algorithmic}[1]
        \Function{$\maxvalid_{\genesis,T}$}{$\overline{\chain}$}
            \Let{\chain_\mmax}{[\,]}
            \Let{\maxwork}{0}
            \For{$\chain \in \overline{\chain}$}
                \If{$\lnot\validate_{\genesis}(\chain)$}
                    \Continue
                \EndIf
                \Let{\thiswork}{\awork(\chain)} \Comment{Computed as $\sum_{\block \in \chain} \hash(\block)_{:\secparam}$}
                \If{$\thiswork > \maxwork$}
                    \Let{\chain_\mmax}{\chain}
                    \Let{\maxwork}{\thiswork}
                \EndIf
            \EndFor
            \State \Return{$\chain_\mmax$}
        \EndFunction
    \end{algorithmic}
\end{algorithm}

\else

\appendix

\section{Chain Selection and Ledger Extraction}
Here, we present two algorithms omitted from the main body: the
ledger-extraction function of $\protocol$ (Algorithm~\ref{alg.extract}), 
and an example instantiation of $\maxvalid$ (Algorithm~\ref{alg.poem-maxvalid}).
We illustrate $\maxvalid$ with PoEM~\cite{poem}, which selects the chain with
the most accumulated intrinsic work. PoEM is the natural pairing because it
uses the same intrinsic-work measure that $\protocol$ uses for reward
weighting; any other host protocol's chain-selection rule applies as well,
including Bitcoin's longest-chain rule.  

\begin{algorithm}[ht]
    \caption{\label{alg.poem-maxvalid} The PoEM maxvalid algorithm}
    \begin{algorithmic}[1]
        \Function{$\maxvalid_{\genesis,T}$}{$\overline{\chain}$}
            \Let{\chain_\mmax}{[\,]}
            \Let{\maxwork}{0}
            \For{$\chain \in \overline{\chain}$}
                \If{$\lnot\validate_{\genesis}(\chain)$}
                    \Continue
                \EndIf
                \Let{\thiswork}{\awork(\chain)} \Comment{Computed as $\sum_{\block \in \chain} \hash(\block)_{:\secparam}$}
                \If{$\thiswork > \maxwork$}
                    \Let{\chain_\mmax}{\chain}
                    \Let{\maxwork}{\thiswork}
                \EndIf
            \EndFor
            \State \Return{$\chain_\mmax$}
        \EndFunction
    \end{algorithmic}
\end{algorithm}

\begin{algorithm}[ht]
    \caption{\label{alg.extract} The ledger extraction mechanism}
    \begin{algorithmic}[1]
        \Function{$\extract_{\safetyParam, \genesis}$}{$\chain$}
            \If{$\lnot \validate_{\genesis}(\chain)$}
                \State \Return $\bot$
            \EndIf
            \Let{\ledger}{[]} \Comment{Initialize empty ledger}
            \For{$\block \in \chain[:-\safetyParam]$} \Comment{Parse each stable block}
                \For{$\tx \in \block.\txList$}
                    \Let{\ledger}{\ledger \concat \tx.\payload}
                \EndFor
            \EndFor
            \State \Return{$\ledger$}
        \EndFunction
    \end{algorithmic}
\end{algorithm}

\section{Extended Analysis}
\label{sec:extended_analysis}
In this section, we provide the remaining parts of the analysis that were omitted from the main body of the paper due to space constraints. The analysis is largely based on FruitChains~\cite{DBLP:conf/podc/PassS17}.

\fi

\fi

\end{document}